\documentclass[fleqn,usenatbib,useAMS]{mnras}

\usepackage{graphicx}	
\usepackage{amsmath}	
\usepackage{amssymb}	
\usepackage{multicol}        
\usepackage{bm}		
\usepackage{pdflscape}	
\usepackage{xcolor}

\usepackage[T1]{fontenc}
\usepackage{ae,aecompl}

\title[Single Transit Candidates from K2]{Single Transit Candidates from K2: Detection and Period Estimation}

\author[H. P. Osborn]{H.P. Osborn$^{1}$\thanks{Contact e-mail: \href{h.p.osborn@warwick.ac.uk}{h.p.osborn@warwick.ac.uk}}, D.J. Armstrong$^{1,2}$, D.J.A. Brown$^{1}$, J. McCormac$^{1}$, A.P. Doyle$^{1}$,
\newauthor T.M. Louden$^{1}$,  J. Kirk$^{1}$, J.J. Spake$^{3}$, K.W.F. Lam$^{1}$, S.R. Walker$^{1}$, F. Faedi$^{1}$,
\newauthor D.L. Pollacco$^{1}$.
\\
$^{1}$University of Warwick, Department of Physics, Gibbet Hill Road, Coventry, CV4 7AL, UK\\
$^{2}$ARC, School of Mathematics \& Physics, Queens University Belfast, University Road, Belfast BT7 1NN, UK\\
$^{3}$Astrophysics Group, School of Physics, University of Exeter, Stocker Road, Exeter EX4 4QL, UK}

\date{Last updated 2015 May 22; in original form 2013 September 5}

\pubyear{2015}

\begin{document}
\label{firstpage}
\pagerange{\pageref{firstpage}--\pageref{lastpage}}
\maketitle

\begin{abstract}
Photometric surveys such as \textit{Kepler} have the precision to identify exoplanet and eclipsing binary candidates from only a single transit. 
K2, with its 75d campaign duration, is ideally suited to detect significant numbers of single-eclipsing objects.
Here we develop a Bayesian transit-fitting tool ("\texttt{Namaste}: An Mcmc Analysis of Single Transit Exoplanets") to extract orbital information from single transit events.
We achieve favourable results testing this technique on known \textit{Kepler} planets, and apply the technique to 7 candidates identified from a targeted search of K2 campaigns 1, 2 and 3.
We find EPIC203311200 to host an excellent exoplanet candidate with a period, assuming zero eccentricity, of $540 ^{+410}_{-230}$ days and a radius of $0.51 \pm 0.05 
R_{\textrm{Jup}}$.
We also find six further transit candidates for which more follow-up is required to determine a planetary origin. 
Such a technique could be used in the future with \textit{TESS}, \textit{PLATO} and ground-based photometric surveys such as NGTS, potentially allowing the detection of planets in reach of confirmation by \textit{Gaia}.
\end{abstract}




\section{Introduction}
Wide-field survey telescopes have for more than a decade searched for the repeated transits of exoplanets, detecting thousands of candidates and confirmed planets.
With photometric precision orders of magnitudes better than on the ground, the \textit{Kepler} mission \citep{borucki2010kepler} has contributed most to this growing field, from sub-Earth-radius worlds \citep{barclay2013sub} to long-period gas giants \citep{wang2013planet}.


In 2014 \textit{Kepler} was repurposed as K2 \citep{howell2014k2} which, due to engineering constraints, observes multiple fields in the ecliptic on 75 day campaigns.
Although reduced pointing stability limits the photometric precision of K2, many stars have been observed with precision on the order of 100ppm per half hour cadence. 
More than 40 planet candidates have so far been detected from the first 3 campaigns of K2 \citep[e.g.][]{foreman2015systematic}, with orbital periods up to 50 days. More than a dozen of these systems have subsequently been validated \citep[e.g.][]{montet2015stellar, armstrong2015one}.

The occurrence rates of transiting planets with periods on the order of months (themselves derived by \textit{Kepler}, e.g. \cite{fressin2013false}) suggest that a handful of longer-period planets should be detected per K2 campaign.
The reduced mission duration of 75 days (compared to 1400 in the primary mission) means that such planets are likely to only transit once.
Planets with transit depths over $\sim1mmag$ offer the potential of discovery in just a single event.
Such signals may then constitute strong planetary candidates with undefined orbital parameters, similar to those detected by microlensing surveys \citep{bennett1996detecting}.


One such planet has previously been detected and confirmed from K2. 
K2-1 b was initially spotted in a single transit during the best 6.5 days of engineering data \citep{vanderburg2015characterizing}.
Subsequent follow-up with both spectroscopy (HARPS) and photometry (MOST) determined that the planet was a $2.5R_{\oplus}$ Earth on a 9.12d orbit.
However, in subsequent campaigns, planets capable of being detected in a single transit are likely to have orbital periods of 40 days or more, making follow-up more challenging.


We show here that for favourable transits, provided the star can be characterised, the orbital period can be estimated from the information contained within a single transit.
We apply this technique to 6 \textit{Kepler} planets as proof of concept and 7 new candidate events detected in K2 data.

\subsection{Single Transit Event Occurrence Rate}
From its inital sampler of $\sim150,000$ stars, \textit{Kepler} detected 101 planets and 828 planet candidates with periods longer than 50 days.
53\% of these exhibit deep enough transits to allow their detection from a single transit ($> 5\sigma$).
We would therefore expect, with $\sim40,000$ stars now observed by K2, to detect substantial numbers of such planets.

A simple analysis, using \textit{Kepler} occurrence rates from 0 to 85 days \citep{fressin2013false} and assuming a flat distribution in $ln(P)$ beyond 85d, suggests that $15\%$ of FGK stars should have a large Neptune or Jupiter on $<3000$ day orbits.
This is similar to the giant planet occurrence rates found by \citep{mayor2011harps}, with $ 14 \pm 2 \% $ of FGK stars with planets larger than $>50M_{\oplus} $. 
Accounting for transit (scaled with $ \frac{R_{\star}}{a} $) and timing probability (scaled with $ \frac{t_{\textrm{obs}}}{P} $) gives detectable multi- and single-transiting planets around $ 0.15\% $ \& $ 0.03\% $ of stars from \cite{fressin2013false} and $ 0.09\% $ \& $ 0.02\% $ from \cite{mayor2011harps}.
Hence one in every few thousand FGK stars observed by K2 should have detectable single transits.

\begin{figure}
	\includegraphics[width=\columnwidth]{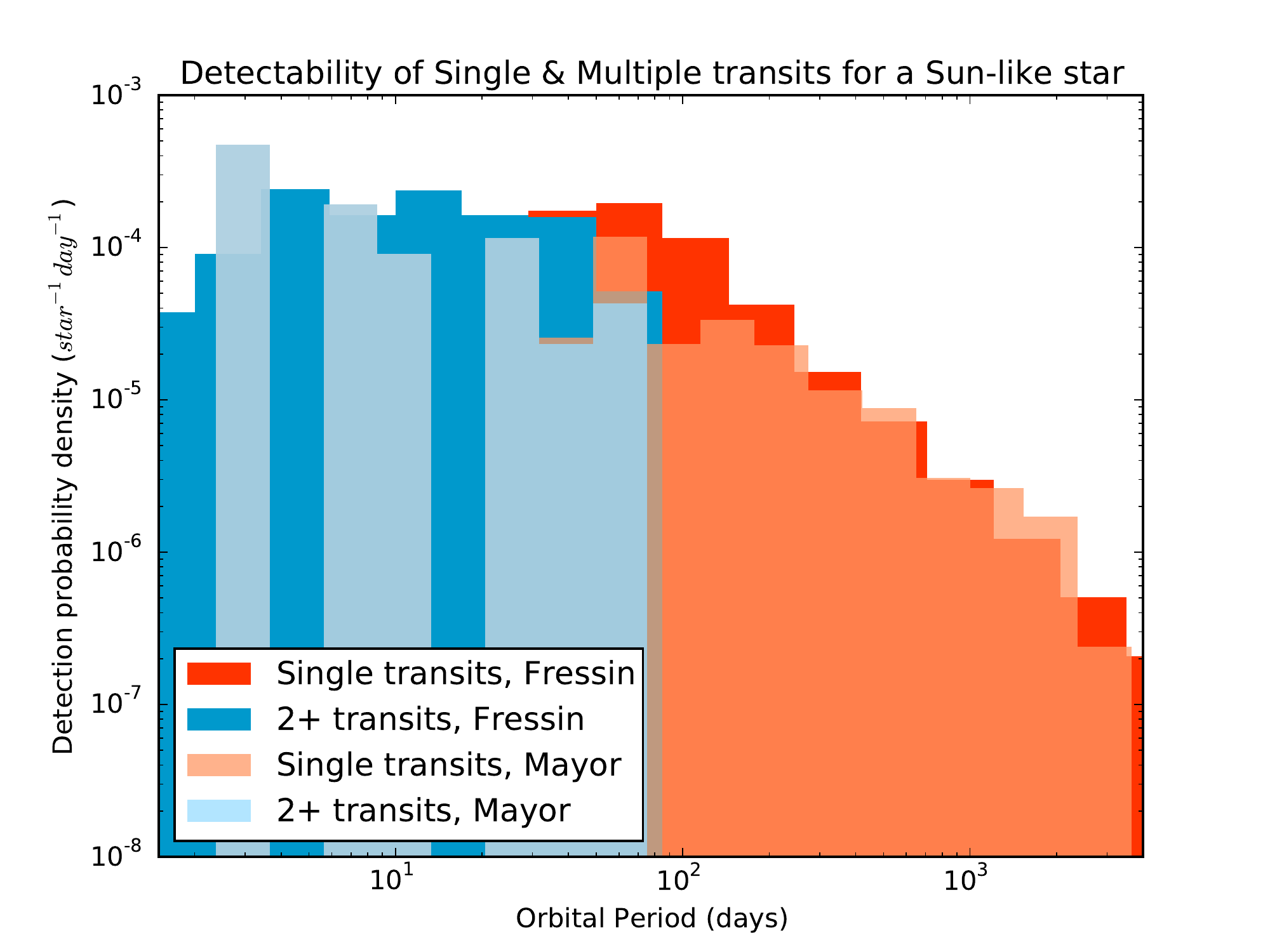}
    \caption{Comparing the detectability of a planet around a sun-like star with K2. Blue: multiple transits detectable. Red: Only single transits detectable}
    \label{fig:eccVdists}
\end{figure}

The decreasing probabilities with orbital distance also suggest that high-precision, shorter-duration surveys such as TESS \citep{ricker2015transiting} or NGTS \citep{wheatley2013next} could detect monotransits in substantial numbers. 
The 28d survey duration for the majority of the TESS field, for example, would likely find monotransiting giant planets around $>0.04\%$ FGK stars, potentially improving giant planet yield by as much as 50\%.

Occurrence rate estimations of this regime are, at present, extremely poorly constrained.
Hence the detection and subsequent follow-up of monotransiting planets in this regime will allow improved occurrence rates for this relatively unexplored parameter space.

\section{Methods}
\subsection{Detrending}
Three campaigns of K2 target pixel files, were obtained from the Mikulski Archive for Space Telescopes (MAST).
Limiting our analysis to objects classified as 'STARS' we performed aperture photometry on the target object.
Aperture sizes were varied according to brightness, with radii of 3, 4, 5 and 12 pixels used for \textit{Kepler} magnitudes bins with boundaries 16, 13 \& 10.
The extracted flux was then background-subtracted, with background RMS added in quadrature to the flux errors. 

To remove the majority of systematics in K2 lightcurves, which are dominated by pointing drift, we independantly developed a detrending method similar to \citep{vanderburg2014technique} to remove all noise correlated with shifting target position regardless of its source.
To do this we compute centroid position for each timestamp, cutting the largest pixel shifts (eg, during thruster firings).
We then create a 2D surface of raw flux against x and y position. 
By binning this to an evenly spaced $10\times10$ grid (discarding bins with fewer than 3 points) and interpolating the median fluxes with SciPy's griddata function \citep{jones2001} we created a smooth surface map of the variation due to centroid shifts.
This could then be divided out of the extracted lightcurve, with the result of significantly decreasing the RMS error of the lightcurve.
Variations not related to pointing drift (eg. long-timescale flux drifts) are not removed by this technique and we note that, in many cases, some systematic noise remains due to other instrumental effects.
This method is described in detail in \cite{armstrong2015k2}.\footnote{Detrended lightcurves are publicly available on MAST at https://archive.stsci.edu/prepds/k2varcat/}.


\subsection{Transit Search}
Long-duration variability was removed from the detrended lightcurves by fitting 3rd order polynomials to 2d windows either side of an untouched 4hr central window.
The fit for each was iterated 20 times, with points $>5\sigma$ from the best fit excluded each time.
This method is further explained in \citep{armstrong2014abundance}
The resulting polynomial fit was then applied to the central 4hr windows, thus avoiding artificially reducing transit depths.

A search for transit signals was then performed on each lightcurve. 
Least-square minimisation was used to fit pre-generated transit models (developed from the \citealt{mandel2002analytic} small planet, quadratic limb-darkened model) to a window of the lightcurve 6 times the transit duration ($T_{D}$).
This was repeated for transit models with durations from 1.5 to 24hrs in increasing $T_{D}$ steps of 25\%, with each fitting window shifted by 25\% of the targeted $T_{D}$ each time.
Well-fitting models with depths greater than $2.5\sigma$ from the out-of-transit RMS were recorded. 
A combination of highest SNR and lowest reduced $ \chi ^2$ value was used to select the best transit fit when multiple durations and transit centres were flagged on the same region of lightcurve.
 
To reduce false positives from thruster firings, the SNR of detections with above-average numbers of events occurring concurrently were suppressed.
Lightcurves were then sorted by the total SNR of detected signals and then manually 'eyeballed' by at least two independent observers, leaving only the best candidate events.
\\
For Campaigns 2 and 3 we limited the search to stars brighter than 13th magnitude, a threshold beyond which RV follow-up is impractical.

\begin{figure}
	\includegraphics[width=1.1\columnwidth]{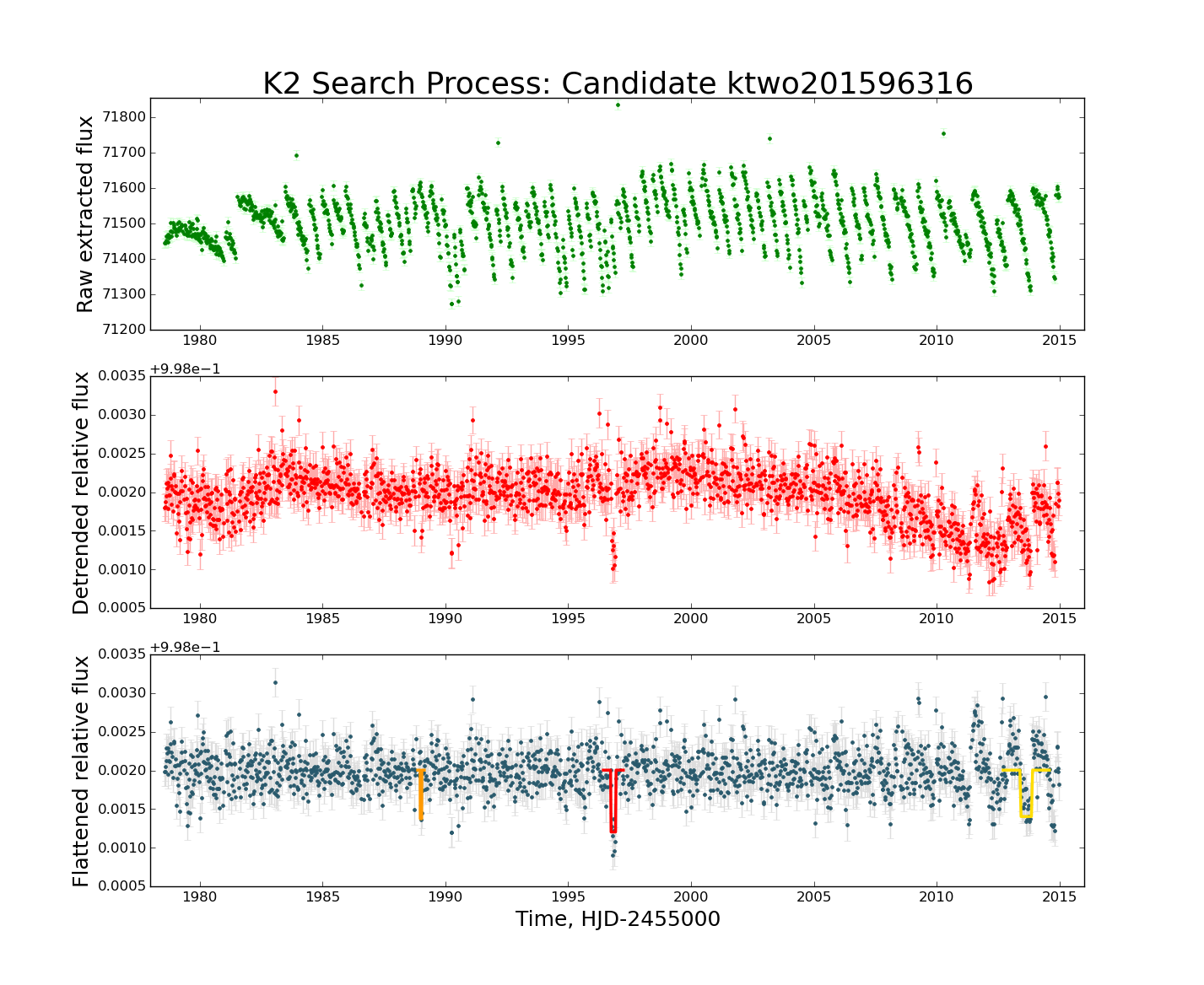}
    \caption{Example of the detection process for validated, multi-transiting planet EPIC201596316 \citep{montet2015stellar}. (a) Raw aperture-extracted flux; (b) Relative flux after de-trending for pixel motion; (c) Transit-searched \& trend-removed with detections in red/orange.}
    \label{fig:eccVdists}
\end{figure}

\subsection{Transit Fitting - \texttt{Namaste}}
Modelling transit lightcurves has been explored by numerous authors \citep[e.g.][]{mandel2002analytic, seager2003unique, cameron2007efficient}, but the majority of full transit models rely on knowledge of the period (often scaled to transit duration) or semi-major axis (scaled with stellar radii).
In the case of a single transit, these approximations cannot be used.
Instead, we develop "\texttt{Namaste}: An Mcmc Analysis of Single Transiting Exoplanets" (hereafter, \texttt{Namaste}\footnote{Publicly available at https://github.com/hposborn/namaste}).
This technique estimates a planetary velocity scaled to stellar radius ($v'$) in place of a velocity calculated from the planetary period.
This velocity can be geometrically defined from impact parameter($b$), planet-to-star ratio (${R_p}/R_{\star}$) and transit duration ($T_{D}$) (equation \ref{eq:vprime})

\begin{equation}
\centering
\label{eq:vprime}
v' \equiv \frac{v_{pl}}{R_{\star}} = \frac{2 \sqrt{(1+R_p/R_{\star})^2 -b^2}}{T_{D}}
\end{equation}

The scaled velocity of a small planet crossing the centre of the stellar disc ($b=0$) is therefore twice the inverse of the transit duration ($\sim 2/T_{D}$).

Velocity, impact parameter \& radii ratio can be estimated geometrically from fitting the transit shape.
We adapt the transit fitting regimes of Ian Crossfield\footnote{Accessed from http://www.lpl.arizona.edu/~ianc/python/} 
 and the Monte Carlo Markov chain implementation \texttt{emcee} \citep{foreman2013emcee}, to estimate posterior probability distributions for each monotransit signal.
Quadratic limb darkening parameters adapted for the \textit{Kepler} bandpass were interpolated from stellar temperature \citep{sing2010stellar}.
A Gaussian prior distribution was applied to these limb darkening parameters with values and errors set from the temperature probability distribution.
Errors were set to the RMS of the 4 different colour estimates to a minimum of 150K.

If eccentricity is assumed to be zero, a circular planetary period ($P_{\textrm{circ}}$) can be estimated from the scaled transit velocity ($v'$) and stellar density ($\rho_{\star}$) using \textit{Kepler}'s laws:

\begin{equation}
\centering
\label{eq:pcirc}
P_{\textrm{circ}}=\frac{8\pi^{2}G}{3} \frac{\rho_{\star}}{v'^{3}} = 2\pi \frac{g}{R_{\star} v'^{3}}
\end{equation}

Longer period orbits (and hence lower-velocity fits) are probabilistically less likely due to transit probability ($p_{\textrm{tr}} \approx \frac{R_{s}}{a_{\textrm{pl}}} \approx \frac{v'}{\rho_{\star}} $).
We discourage longer-period fits (and encourage faster-velocity fits) with a linear prior on transit velocity.
In the case of multi-planet systems, the probability of a further planet transiting does not simply scale with transit probability $R_{\star}/a$, as co-planar orbits are favoured. 
Hence, in these cases, the forcing of fits to shorter orbits by a linear prior may not be valid.
However, the increase in transit probability of a planet at distance $x$ given transiting exoplanets on orbits $y$,$z$,etc is a complex problem beyond the scope of this work.
The non-detection of subsequent transits in the lightcurve can also be used to set a lower limit on the orbital period, and hence an upper limit on velocity.
We do not apply this technique in order to produce transit fits fully independent of stellar parameters (e.g. density).

The impact parameter was limited to the range -1.2 to 1.2 to avoid walkers building up at $b=0$. 
Planet-to-star radius ratios were limited to 0.25, as above this the assumption that the transiting object is fully opaque and covering a uniform region of stellar surface breaks down.

Velocity, just like the transit duration scaled to period, is directly linked to stellar density \citep{seager2003unique}.
For the best constrained transit models, stellar density is likely to prove the largest uncertainty.
Characterising the star, therefore, is key to estimate orbital period. 
Such characterisation is best performed with asteroseimology, or less accurately, with spectral fitting.

\begin{figure}
	\includegraphics[width=\columnwidth]{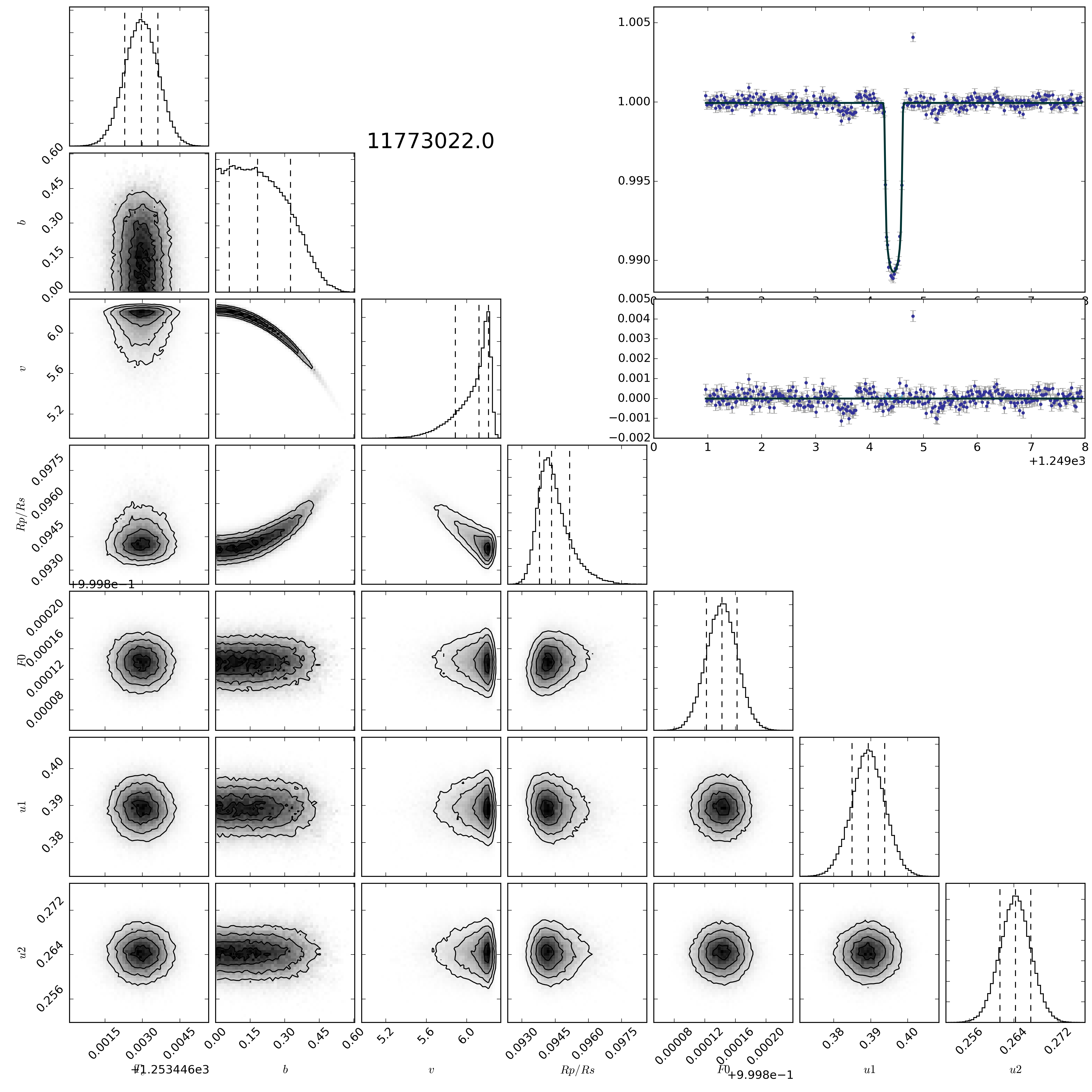}
    \caption{An example of the posterior distribution produced by \texttt{Namaste}. In this case for Kepler-51 d (KIC11773022). Triangle plots from all fits can be found in Appendix A.}
    \label{fig:NamastePostEg}
\end{figure}

Large radii planets which spend longer crossing the rim of the stellar disc are most suitable to \texttt{Namaste} fitting as the impact parameter can be more easily distinguished.
\cite{price2015low} showed that for smaller and lower signal-to-noise planet transits the uncertainty on impact parameter increases linearly, causing poor determination of perpendicular velocity and therefore eccentricity.
For less well-defined fits, parameters such as impact parameter, planet-to-star ratio and velocity become correlated, as can be seen in their posterior distributions of figure \ref{fig:NamastePostEg}.
This is especially true for eclipses that cannot be constrained to an impact parameter less than 1.0.
In these cases the fit cannot distinguish between planetary, high-velocity disc-crossing transits and lower-velocity grazing eclipses.
As impact parameter increases beyond 1.0, velocity stabilises to a minimum value determined by transit duration.
However, even for correlated and non-gaussian parameters, \texttt{Namaste} allows us to put probabilistic constraints on the transit fit.

An example of the resulting posterior distributions between parameters (along with model fit and residuals) can be seen in figure \ref{fig:NamastePostEg}

\subsubsection{Eccentricity}
For exoplanets on non-circular orbits, the circular velocity estimated by \texttt{Namaste} ($v_{\textrm{circ}}$) depends strongly on eccentricity and the argument of periastron (equation \ref{vecc}, \citealt{barnes2007effects}).

\begin{equation} \label{vecc}
V_{\theta} = V_{\textrm{circ}}\frac{1 + e \cos \theta}{\sqrt{1 - e^2}}
\end{equation}

\begin{figure}
	\includegraphics[width=\columnwidth]{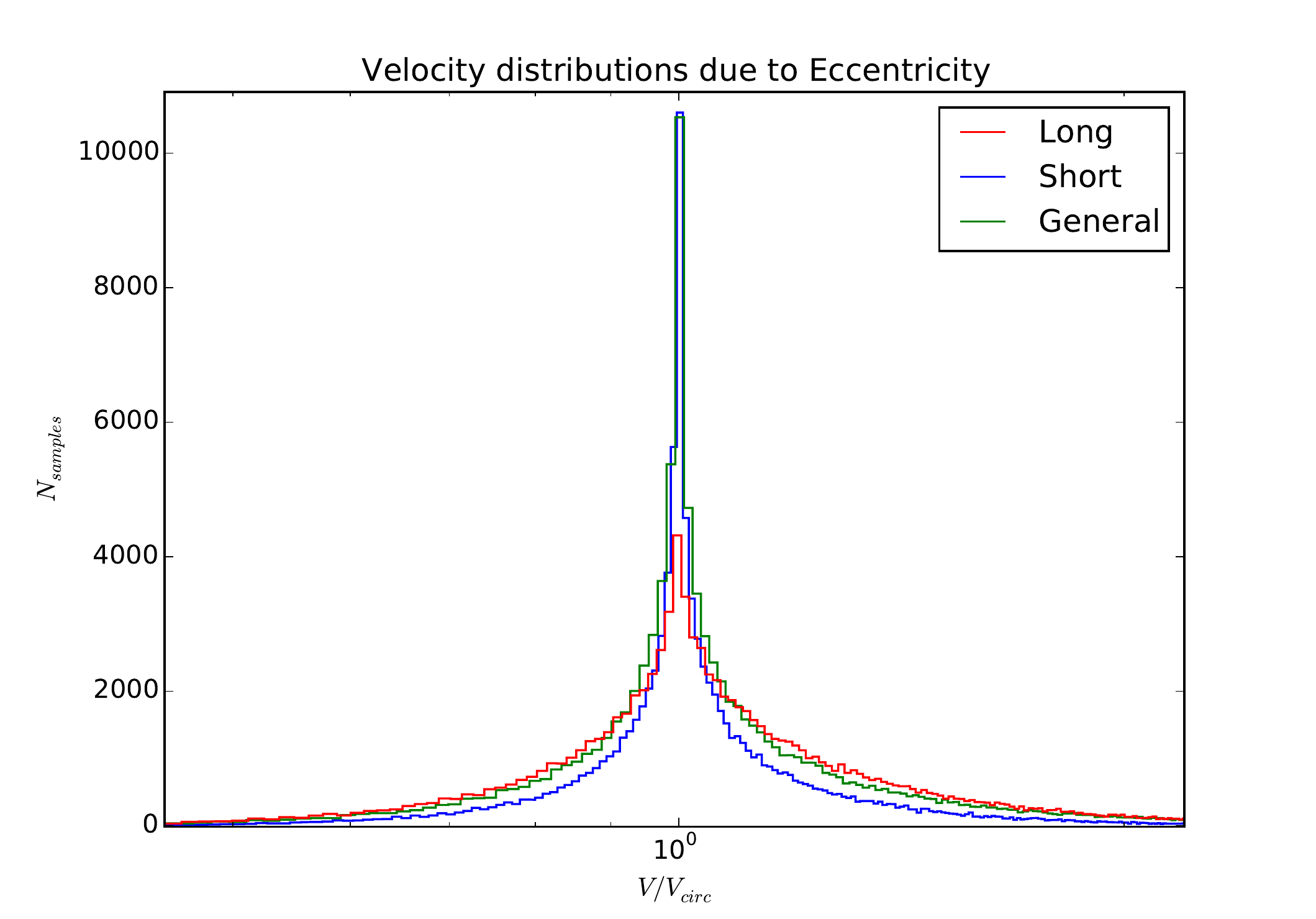}
    \caption{Distribution of the ratio of true velocity and circular velocity for different three eccentricity distributions from \citep{kipping2013parametrizing}.}
    \label{eccVdists}
\end{figure}

As the solid angle swept out by an eccentric planet's shadow is greater than that from an equivalent circular orbit, eccentric planets are also more likely to transit.
Transit probability is especially raised near periapse, suggesting circular period estimates will on average underestimate the true period.
\cite{kipping2013parametrizing} used RV planets to study the distribution of exoplanet eccentricities and showed that close-in planets (defined as $P<382d$) have a more circular distribution than long-period planets.
We use these distributions to study eccentricity's effect on transit velocity (Figure \ref{eccVdists}).
Confidence intervals.
We find that eccentricity increases the median velocity and its $1-\sigma$ confidence intervals by $1.3^{+21}_{-7} \% $ in the short case, and by $3^{+35}_{-9}\% $ in the general case.

\cite{van2014asteroseismology} studied the discrepancy between \textit{Kepler} planet densities determined by asteroseismology and those found from transit duration to determine their eccentricities.
This method is analogous to the comparison of true period with those from single transit fits, and suggests \texttt{Namaste} could be useful for determining the eccentricities of polytransiting planets.

In this study, we note that eccentricity can add significant uncertainty to our results, but limit ourselves to estimations of circular periods ($P_{\textrm{circ}}$) which are good approximations for the majority of cases.
For the short-period regime, two-thirds of planets orbit with eccentricities less than 0.2, causing substantially lower increased uncertainties.
Hence, in the majority of cases, the small increase in velocity uncertainty is negligible compared to the large uncertainties from stellar density.


\subsection{Stellar Parameter Fits}
Where stellar parameters from spectra are unavailable, we use photometric colours to approximate the temperatures, masses and radii of EPIC stars.
The majority of K2 stars have broadband photometric measurements in both the visible (e.g. Tycho B \& V bands from \cite{hog2000tycho}) and infra-red (e.g. 2MASS J,H,K data from \cite{cutri20032mass}).
From the four independant colours derived from these magnitudes, stellar temperatures can be estimated \citep{fitzgerald1970intrinsic}.
This relationship assumes that no bright companions are present, which can only be confirmed with more advanced follow-up.
From this temperature, and making the assumption that the star is on the main sequence, stellar models allow a stellar radius \& mass to be estimated \citep{torres2010accurate}. 
Temperature uncertainty was estimated from the RMS of the temperature from each colour estimate, or 150K if the RMS was smaller than this value.
Mass and radius were taken from these temperature uncertainties down to a minimum relative error of 10\%.




\section{Results}

\subsection{Application to Known \textit{Kepler} Systems}
\begin{figure}
	\centering
	\includegraphics[width=\textwidth]{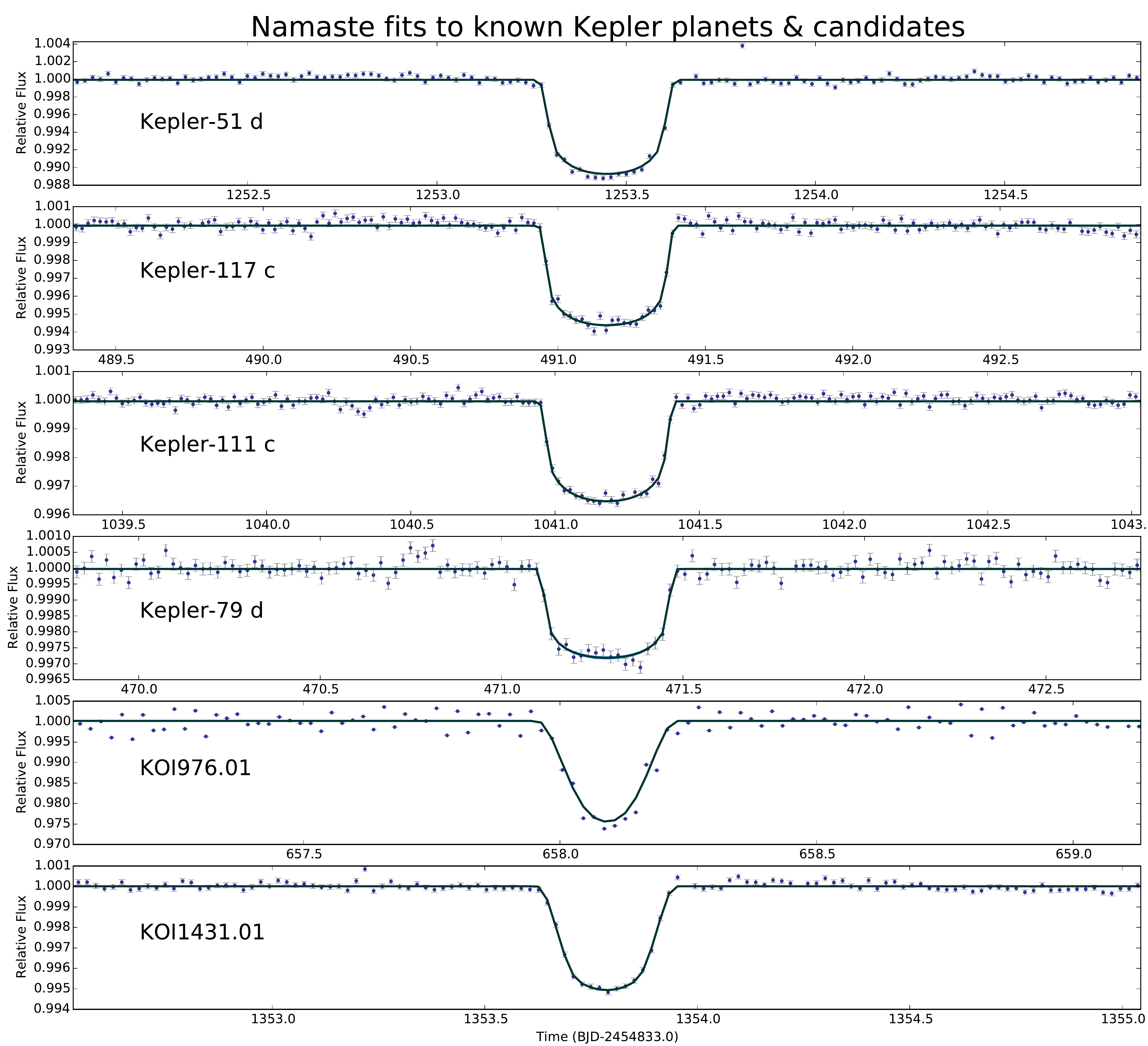}
    \caption{Six long-period \textit{Kepler} planets with \texttt{Namaste} fits. Best-fit models are in black, while $1-\sigma$ error regions are in blue. x-axis is scaled to 8 transit durations, whereas the y-axis is unconstrained.}
    \label{fig:KepFits}
\end{figure}

To test the fitting of \texttt{Namaste}, we applied it to single transits from the lightcurves of 6 long-period \textit{Kepler} planets and KOI candidates.
Figure \ref{fig:KepFits} shows the \texttt{Namaste} model fits to these transits and Table \ref{KepFitsTable} gives the output parameters compared to their published values.

\onecolumn

\begin{table}
\centering
\label{KepFitsTable}
\resizebox{\textwidth}{!}{%
\begin{tabular}{lllllll}
\hline
& \textbf{Kepler-51 d} & \textbf{Kepler-117 c} & \textbf{Kepler-111 c} & \textbf{Kepler-79 d} & \textbf{KOI976.01} & \textbf{KOI1431.01} \\
\hline
KIC & 11773022 & 10723750 & 8559644 & 8394721 & 3441784 & 11075279 \\[2pt]
$T_{\textrm{cen}} (d)$     & $1253.4490 \pm 0.0006$        & $491.16605 \pm 0.0009$           & $1041.17996 \pm 9\times 10^{-4}$ & $471.290 \pm 0.001$                     & $658.09145 \pm 2 \times 10^{-5}$   & $1353.7881 \pm 6.6\time 10^{-4}$ \\[2pt]
{\color{black!60}$T_{\textrm{cen,true}}$} & {\color{black!60}$1253.454 \pm 0.05$} & {\color{black!60}$491.164 \pm 0.03$} & {\color{black!60}$1041.210 \pm 0.07$} & {\color{black!60}$471.290 \pm0.04$} & {\color{black!60}$658.093 \pm 0.05$} & {\color{black!60}$1353.788 \pm 0.06$} \\[2pt]
$b$               & $0.18^{+0.13}_{-0.11}$        & $0.32^{+0.12}_{-0.19}$           & $0.56^{+0.10}_{-0.29}$           & $0.43 \pm0.23$                          & $0.8420 \pm 7 \times 10^{-4}$      & $0.872 \pm 0.006$               \\[2pt]
{\color{black!60}$b_{true}$} & {\color{black!60}$0.094 \pm 0.08$} & {\color{black!60}$0.20^{+0.05}_{-0.14}$} & {\color{black!60}$0.787^{+0.007}_{-0.02}$} & {\color{black!60} $0.02^{+0.15}_{-0.02}$} & {\color{black!60}$1.29^{+0.34}_{-0.17}$} & {\color{black!60}$0.869^{+0.005}_{-0.007}$} \\[2pt]
$v (R_{\star} d^-1)$    & $6.1215^{+0.090}_{-0.21}$     & $4.51^{+0.20}_{-0.25}$           & $3.85^{+0.48}_{-0.34}$           & $5.2^{+0.4}_{-0.9}$                     & $6.191 \pm 0.006$                  & $4.03^{+0.10}_{-0.09}$           \\[2pt]
$R_p/R_{\star}$              & $0.0943^{+0.0007}_{-0.0005}$  & $0.0690 \pm 6\times 10^{-4}$     & $0.0560 \pm 0.001$              & $0.049^{+0.002}_{-0.001}$               & $0.165 \pm0.0002$                & $0.0759 \pm 7 \times 10^{-4}$    \\[2pt]
{\color{black!60}$R_p/R_{\star,\textrm{true}}$} & {\color{black!60} $0.11^{+0.015}_{-0.08}$} & {\color{black!60} $0.057^{+0.03}_{-0.015}$ } & {\color{black!60} $0.06^{+0.03}_{-0.02}$} & {\color{black!60}$0.0372^{+0.02}_{-0.005}$ } & {\color{black!60}$0.5^{+0.1}_{-0.4}$} & {\color{black!60}$0.076^{+0.04}_{-0.03}$} \\[2pt]
$F$               & $0.99994 \pm 2\times 10^{-5}$ & $0.99994 \pm 1.3\times 10^{-5} $ & $0.999961 7\pm \times 10^{-6}$   & $0.99997 \pm 1.0\time 10^{-5}$          & $1.0001733 \pm 1.3 \times 10^{-6}$ & $0.999958 \pm 8 \times 10^{-6}$  \\[2pt]
$u1$              & $0.389 \pm 0.004$             & $0.33^{+0.003}_{-0.004}$         & $0.363 \pm 0.006$                & $0.327 \pm 0.003$                      & $0.2616 \pm 0.003$                 & $0.421 \pm 0.006$               \\[2pt]
$u2$              & $0.264 \pm 0.0025$            & $0.297 \pm 0.002$               & $0.280 \pm 0.003$               & $0.298^{+0.002}_{-0.003}$               & $0.3468 \pm 0.0015$                & $0.245 \pm 0.004$                \\[2pt]
$R_P (R_{\textrm{Jup}})$   & $0.84 \pm 0.41$               & $0.99 \pm 0.17$                  & $0.63 \pm 0.07$                  & $0.7 \pm0.1$                            & $2.7 \pm 1.1$                     & $0.70 \pm 0.06$                  \\[2pt]
{\color{black!60}$R_{p,\textrm{true}}$} & {\color{black!60}$0.96^{+0.39}_{-0.09}$ } & {\color{black!60} $0.8^{+0.5}_{-0.1}$ } & {\color{black!60} $0.67^{+0.15}_{-0.07}$ } & {\color{black!60} $0.51^{+0.2}_{-0.05}$ } & {\color{black!60}$8^{+7}_{-1}$ } & {\color{black!60} $0.71^{+0.11}_{-0.03}$} \\[2pt]
$P (d)$           & $138^{+20}_{-10}$             & $53.0^{+5.3}_{-18.7}$            & $240^{+130}_{-90}$               & $55.3 ^{+71}_{-3.1}$                    & $25^{+100}_{-17}$                  & $340^{+110}_{-80}$               \\[2pt]
{\color{black!60}$P_{\textrm{true}}$} & {\color{black!60}$130.1775 \pm 0.0001$} & {\color{black!60}$50.79035 \pm 2\times 10^{-5}$} & {\color{black!60} $224.7782 \pm0.0003$} & {\color{black!60} $52.09082 \pm 4\times 10^{-5}$ } & {\color{black!60}$52.56899 \pm 6 \times 10^{-5}$ } & {\color{black!60}$345.1599 \pm 4 \times 10^{-4}$} \\[2pt]
$s.m.a. (\textrm{AU})$     & $0.5^{+1.4}_{-0.3}$           & $0.36^{+0.17}_{-0.10}$           & $0.80^{+0.27}_{-0.21}$           & $0.3^{+0.2}_{-0.1}$                     & $0.20^{+0.42}_{-0.11}$             & $0.98^{+0.20}_{-0.17}$           \\[2pt]
{\color{black!60}$s.m.a._{\textrm{true}}$} & {\color{black!60}$0.5121$} & {\color{black!60}$0.2858$ } & {\color{black!60}$0.7469$ } & {\color{black!60} $0.2807$ } & {\color{black!60} $0.321$ } & {\color{black!60}$0.9809$} \\        \\[2pt]
$T_S (K)$         & $5800 \pm110$                 & $6170\pm100$                     & $5952 \pm 74$                    & $6174 ^{+83}_{-117}$ & $7200^{+310}_{-240}$               & $5600^{+110}_{-90}$              \\[2pt]
$R_{star} (R_{\odot})$ & $0.91 \pm0.5$                 & $1.61\pm0.05$                    & $1.16\pm 0.14$                   & $1.302^{+0.026}_{-0.027}$               & $1.66^{+0.2}_{-1.4}$               & $0.95^{+0.05}_{-0.14}$           \\[2pt]
$M_{\star} (M_{\odot})$     & $1.04 \pm0.12$                & $1.13^{+0.13}_{-0.02}$           & $1.17 \pm 0.03$                  & $1.165+\{0.044\}_\{-0.045\}$              & $1.59^{+0.18}_{-0.4}$              & $1.054^{+0.08}_{-0.06}$          \\[2pt]
$\rho_S (\rho_{\odot})$ & $1.72 \pm0.12$          & $0.29^{+0.01}_{-0.02}$           & $0.75 \pm 0.16$                  & $0.526^{+0.018}_{-0.024}$               & $0.35^{+0.29}_{-0.30}$                     & $1.22^{+0.27}_{-0.47}$        \\[2pt]
\hline
\end{tabular}
}
\caption{\texttt{Namaste} results for 4 \textit{Kepler} planets and 2 KOI candidates, compared with published parameters from traditional analyses. See Appendix A1 for posterior distributions.}

\end{table}

\subsection{Application to K2 Single Transit Candidates}
We applied \texttt{Namaste} to seven potential single transit events found in K2. Fits are shown in Figure \ref{fig:K2Fits}, fit parameters in table \ref{k2fitstable} and full posterior distributions in Appendix A2.

\begin{figure}
\centering
	\includegraphics[width=\columnwidth]{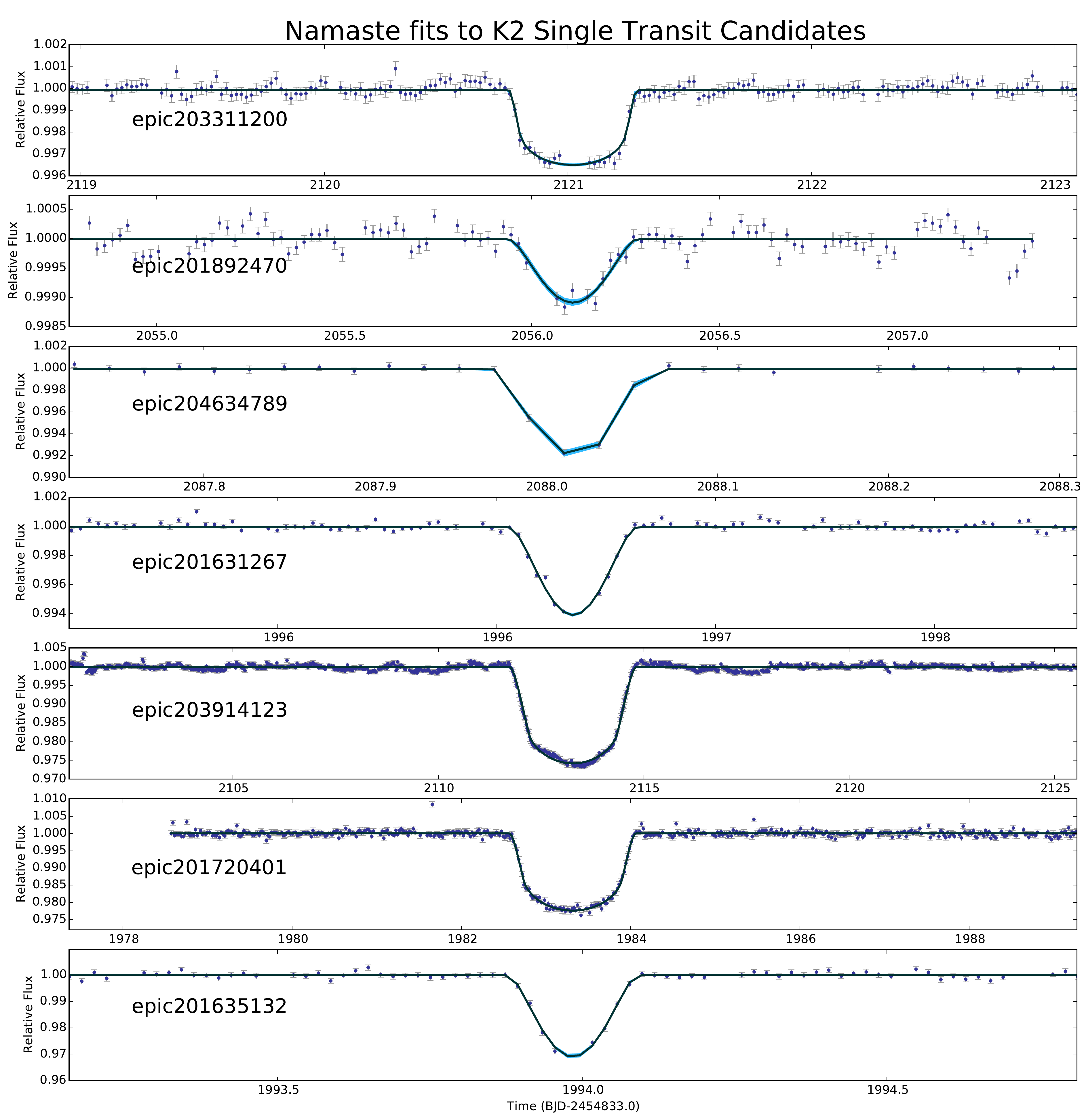}
    \caption{Six new K2 planet-like signals with \texttt{Namaste} fits. Best-fit models are in black with $1-\sigma$ error regions in light blue. x-axis is scaled to 8 transit durations, whereas the y-axis is unconstrained.}
    \label{fig:K2Fits}
\end{figure}

\begin{table}
\centering
\label{k2fitstable}
\resizebox{\textwidth}{!}{%
\begin{tabular}{llllllll}
\hline
 &  \textbf{EPIC203311200} &  \textbf{EPIC201892470} &  \textbf{EPIC204634789} &  \textbf{EPIC201631267} &  \textbf{EPIC203914123} &  \textbf{EPIC201720401} &  \textbf{EPIC201635132} \\
\hline
$KepMag$          & $11.896$                      & $11.587$                   & $12.292$                        & $12.78$                    & $9.039$                    & $14.74$                  & $15.143$                   \\[2pt]
$T_{\textrm{cen}} (d)$     & $2121.021 ^{+0.003}_{-0.002}$ & $2056.11 \pm 0.005 $       & $2088.016 \pm 0.001 $           & $1996.674 \pm 0.001 $      & $2113.278 \pm 0.002 $      & $1983.314 \pm 0.002 $    & $1993.985 \pm 0.001 $      \\[2pt]
$b$               & $0.61 ^{+0.13}_{-0.3}$        & $1.07 \pm 0.08 $           & $0.74 ^{+0.25}_{-0.43}$         & $1.05 ^{+0.09}_{-0.08}$    & $0.49 \pm 0.02 $           & $0.15 ^{+0.1}_{-0.09}$   & $0.97 ^{+0.05}_{-0.07}$    \\[2pt]
$v (R_{\star} d^-1)$    & $3.3 ^{+0.7}_{-0.5}$          & $1.9 ^{+0.2}_{-0.1}$       & $21.7 \pm 8.1 $                 & $3.7 ^{+0.2}_{-0.1}$       & $0.679 ^{+0.006}_{-0.007}$ & $1.51 ^{+0.01}_{-0.03}$  & $7.5 ^{+0.3}_{-0.2}$       \\[2pt]
$R_p/R_{\star}$              & $0.056 ^{+0.003}_{-0.002}$    & $0.121 ^{+0.073}_{-0.062}$ & $0.088 ^{0.038}_{-0.008}$       & $0.176 ^{+0.074}_{-0.057}$ & $0.149 \pm 0.001 $         & $0.133 \pm 0.001 $       & $0.24 \pm 0.038 $          \\[2pt]
$F$               & $0.999947 \pm 1.5\times 10^{-5} $      & $0.999998 \pm 1 \times 10^{-5} $    & $0.999941 \pm 5 \times 10^{-5}$ & $0.999965 \pm 1.3\times 10^{-5} $   & $0.999895 \pm 1.7\times 10^{-5} $   & $1.000087 \pm 3\times10^{-5} $  & $0.999996 \pm 9.0\times 10^{-5} $   \\[2pt]
$u1$              & $0.499 \pm 0.009 $            & $0.524 \pm 0.015 $         & $0.603 \pm 0.002 $              & $0.627 \pm 0.007 $         & $0.591 ^{+0.023}_{-0.028}$ & $0.589 \pm 0.016 $       & $0.039 ^{+0.038}_{-0.026}$ \\[2pt]
$u2$              & $0.193 \pm 0.006 $            & $0.175 \pm 0.011 $         & $0.117 \pm 0.002 $              & $0.099 \pm 0.005 $         & $0.036 ^{+0.047}_{-0.025}$ & $0.119 \pm 0.014 $       & $0.028 ^{+0.027}_{-0.018}$ \\[2pt]
$R_P (R_{\textrm{Jup}})$   & $0.51 \pm 0.05 $              & $1.09 ^{+0.65}_{-0.56}$    & $0.75 ^{0.32}_{-0.1}$           & $1.45 ^{+0.61}_{-0.47}$    & $1.07 \pm 0.1 $            & $1.14 \pm 0.1 $          & $2.03 ^{+0.83}_{-0.76}$    \\[2pt]
$P (d)$           & $540 ^{+410}_{-230}$          & $2790 ^{+1270}_{-970}$     & $2.2 ^{10}_{-1.4}$              & $440 ^{+160}_{-120}$       & $83830 ^{+31600}_{-22320}$ & $6340 ^{+2230}_{-1530}$  & $30 ^{+70}_{-20}$          \\[2pt]
$s.m.a. (\textrm{AU})$     & $1.25 ^{+0.59}_{-0.39}$       & $3.71 ^{+1.09}_{-0.94}$    & $0.03 ^{0.04}_{-0.01}$          & $1.03 ^{+0.25}_{-0.2}$     & $31.17 ^{+8.35}_{-6.71}$   & $6.17 ^{+1.46}_{-1.1}$   & $0.13 ^{+0.19}_{-0.06}$    \\[2pt]
$T_S (K)$         & $5200 \pm 200 $               & $5100 \pm 200 $            & $4800 \pm 200 $                 & $4600 \pm 200 $            & $4000 \pm 400 $            & $4800 \pm 200 $          & $3400 \pm 200 $            \\[2pt]
$R_{\star} (R_{\odot})$ & $0.94 \pm 0.09 $              & $0.93 \pm 0.09 $           & $0.87 \pm 0.09 $                & $0.85 \pm 0.08 $           & $0.74 \pm 0.07 $           & $0.88 \pm 0.09 $         & $0.89 \pm 0.36 $           \\[2pt]
$M_{\star} (M_{\odot})$ & $0.91 \pm 0.09 $              & $0.88 \pm 0.09 $           & $0.78 \pm 0.08 $                & $0.74 \pm 0.07 $           & $0.57 \pm 0.1 $            & $0.78 \pm 0.08 $         & $0.43 \pm 0.04 $           \\[2pt]
\hline
\end{tabular}
}
\caption{Output median parameters for \texttt{Namaste} runs of 7 K2 single transit candidates, with 16\% and 84\% percentile bounds. See Appendix A2 for posterior distributions.}
\end{table}
\twocolumn

\section{Discussion}

\subsection{Known Kepler Planets \& Candidates}
\begin{itemize}
\item \textbf{Kepler-51 d -} 

Planetary velocity, planet-to-star ratio and impact parameter are all well-constrained by this fit. 
Well-derived stellar density from \cite{masuda2014very} also gives period errors on the order of only 10\%, and within $1\sigma$ of the true period of 130.2d.
The small discrepancy between true and estimated periods is likely from the overestimation of impact parameter ($b_{\textrm{est}}=0.18 ; b_{\textrm{true}}=0.094$).
This further suggests Kepler-51 d is on a circular orbit.

\item \textbf{Kepler-117 c -}

\texttt{Namaste} fits show good agreement to the published parameters, with all parameters within errorbars. 
Well-constrained stellar parameters from \cite{bruno2015sophie} produce relatively small errors on the estimated period of 53d, which is within 5\% of the true period of 50.7d.

\item \textbf{Kepler-111 c -}

This multiplanet system, validated by \citep{rowe2014validation}, has less well-constrained parameters than the other \textit{Kepler} test cases, with a 20\% density error.
This gives wide errors on the estimated period of $240^{+130}_{-90}$, although the true period of 224.8d sits well within this distribution, and suggests density uncertainties may be over-estimated.

\item \textbf{Kepler-79 d - }
An ultra-low density gas giant, Kepler-79 is a well-constrained system with masses from TTVs \citep{jontof2014kepler}.
In this case, impact parameter is inaccurately fitted, leading to an overestimated period greater than $1\sigma$ from the true value.
The non-zero eccentricity suggested for this planet may also cause increased discrepancy.

\item \textbf{KOI976.01 - }

Despite being on the \textit{Kepler} candidate list, KOI976 is an eclipsing binary with a much larger radius than would be expected from an exoplanet ($8^{+7}_{-1} R_{\textrm{Jup}}$).
However, this is a good proof of concept for this technique on low-mass eclipsing stellar (or brown dwarf) companions, for which the period estimates are still valid.
The long duration of the eclipse gives extremely well-fitting parameters, with output parameters such as velocity accurate to one part in 1000. 
However these disagree with the published KOI catalogue values in many cases, especially impact parameter, which may be due to the eclipsing body being self-luminous.
This leads to an over-estimated velocity and underestimated period of $25^{+100}_{-17}$ days, although the true period is well within the errorbars.

\item \textbf{KOI1431.01} 

This candidate shows the most gaussian and closest period approximation of all the fits, with a median period within 2\% of the true value.
However, poor (and likely over-estimated) stellar density constraints give large errorbars of $\sim \pm 100d$

\end{itemize}

\subsection{Testing with published \textit{Kepler} planet parameters}
Initially, we used transit data for 102 confirmed \textit{Kepler} planets with stellar radius and mass measurements available from the NASA Exoplanet Archive \citep{akeson2013nasa}\footnote{http://exoplanetarchive.ipac.caltech.edu/}.
From transit duration, impact parameter, planetary radii and stellar radii, a circular transit period was estimated and compared to the known period. 
In the case of eccentric planets, as was detected by \citep{van2014asteroseismology}, the period estimate is likely to be significantly offset from the circular period.

Intriguingly, stellar densities derived from stellar surface gravities for a wider range (660) of \textit{Kepler} planets showed a correlation between estimated period and impact parameter for grazing transits ($b>0.6$).
This effect is likely due to overestimation of the duration and impact parameter of the shortest transits, for example due to smeared TTVs, which would cause an increase in the estimated period.
Such effects would only be present for phase-folded transits, hence unlikely to present problems for \texttt{Namaste} analysis.

\begin{figure}
	\includegraphics[width=\columnwidth]{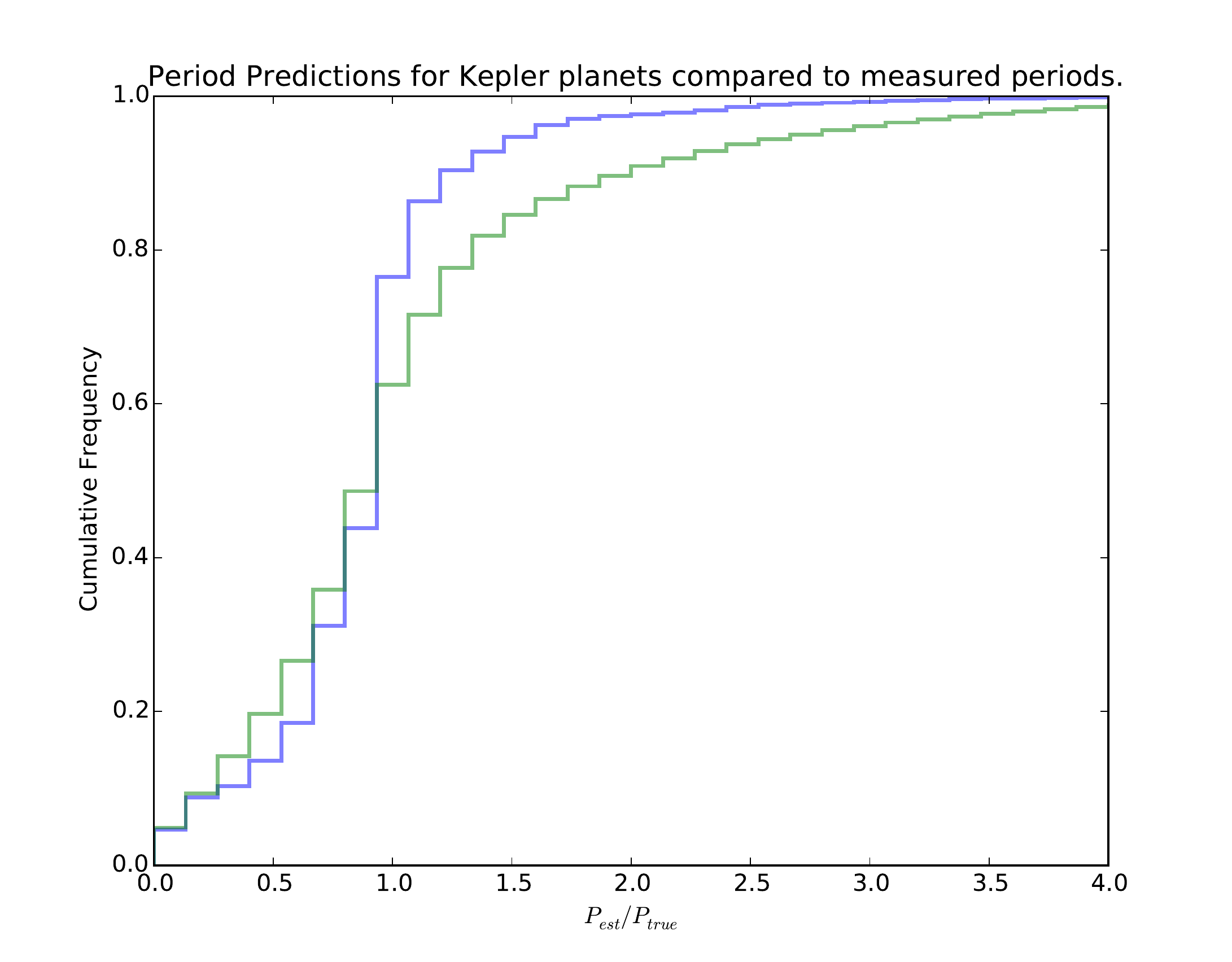}
    \caption{Comparison of real period with true period for two density distributions: $M_{\star}$ \& $R_{\star}$ estimates (green) and directly quotes density values.}
    \label{fig:KepFitsAll}
\end{figure}

To avoid this issue, we compared only the better constrained densities from $R_{\star}$ and $M_{\star}$ estimates, or directly from asteroseismology.
Parameters from the 101 and 69 planets, respectively, were resampled 500 times from a Gassian distribution derived from their uncertainties to produce a cumulative probability distribution of the estimated period compared to the true period can be seen in figure \ref{fig:KepFitsAll}.

As can be seen, the less-precise density estimates from $R_{\star}$ \& $M_{\star}$ give a shallower distribution of orbital period estimates.
However both distributions peak at the true value, and suggest that the analytical technique is valid. 

\subsection{K2 Single Transit Events} 

\subsubsection{EPIC203311200}
\begin{figure}
\centering
	\includegraphics[width=\columnwidth]{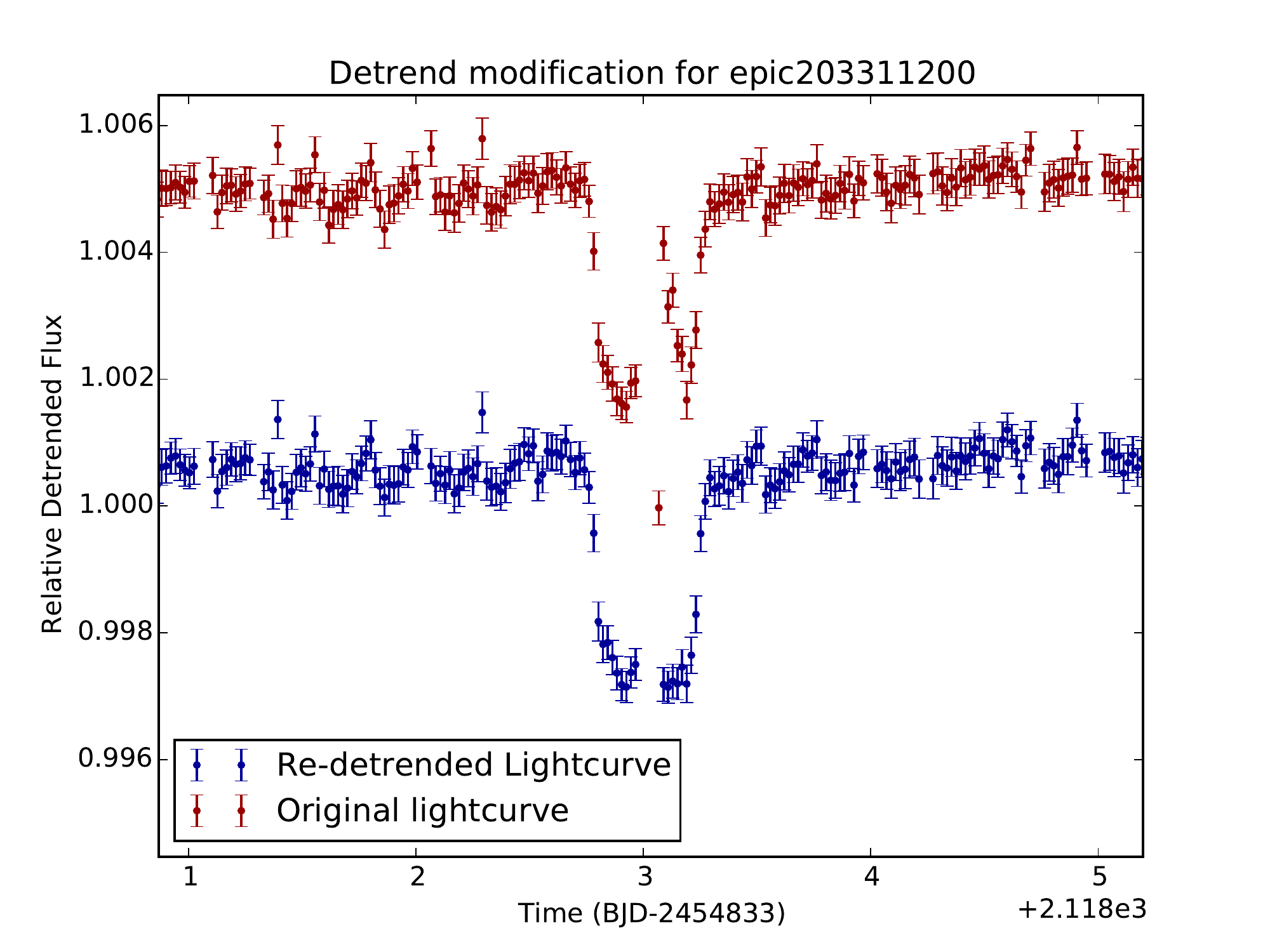}
    \caption{Change in lightcurve of EPIC203311200 after modifying detrending procedure.}
    \label{DetrendMod}
\end{figure}

Unfortunately the centre of this single transit, which is the most obvious feature of the second half of the K2 lightcurve, is coincident with a global position shift at MJD 2121.05 of 1.23 pixels.
This extreme pixel shift was poorly fitted by the interpolated flux map, causing a 5-point bump in the data during the transit (see Figure \ref{DetrendMod}).
To fix the poor detrending for these points, we extrapolated the well-fitted region of 3D fluxmap surface to this region of centroid shift and re-ran the detrending. 
This had the effect of reducing systematic noise across the lightcurve, including those in-transit.
However, six of the largest pixel shifted points were still removed from central transit by this detrending procedure.

The \texttt{Namaste} fit for EPIC203311200 constrains the transit to being a disc-crossing (ie, non-grazing) eclipse, and can therefore put good constraints on the size of the eclipsing body.
It estimates a size of $0.51 \pm 0.05 R_{\textrm{Jup}}$, putting it between the sizes of Uranus and Saturn.
The posteriors show a two-peaked distribution in velocity and radius ratio, which may be due to a build up of walkers at the maximum velocity threshold, or an excess of distribution at $b\sim0$.
We cannot currently separate the correct distribution, although transit probability favours the maximum velocity (with $v'\sim 4 R_{\star}d^{-1}$ giving a potentially more reasonable period of $P_{\textrm{circ}}\sim310d$).
Either period would make EPIC203311200 one of the longest-period transiting exoplanets yet discovered, with the potential to exceed Kepler-421 b ($704.2d$, \cite{kipping2014discovery})

V-J, V-H and V-K show good agreement for a $5200\pm200K$ star, however $(B-V)=0.19$ appears anomalously blue.
A low-resolution spectrum taken with IDS on the 2.5m Isaac Newton Telescope suggests it to be a relatively metal-poor star ($[Fe/H]\sim-1.0$) with a temperature around $5400\pm300 K$; consistent with IR colours.
The low-resolution of the spectrum did not enable a log-g estimate.
If no source of dilution is present, and the star is a main sequence late-G star as suggested by photometric colours and spectra, then the source of this 3mmag eclipse signal must be planetary.

\subsubsection{EPIC201892470}
This shallow but V-shaped eclipse occurs only 1.2d before the end of the K2 lightcurve.
This allows us to place a high lower limit ($>77d$) on the orbital period.
A grazing eclipse is favoured by \texttt{Namaste} (b=1.07).
The long circular period of $2790 ^{+1270}_{-970}$ estimated by \texttt{Namaste} is likely unphysical, and could be an indication that this is a giant rather than a main sequence star.
However it should be noted that the more planetary posterior regions - those with smallest $R_p/R_{\star}$, lowest $b$ and hence larger $v'$ - give the shortest and therefore more likely periods.

\subsubsection{EPIC204634789}
With only four in-transit datapoints, the fit is poorly constrained and is unable to distinguish between any impact parameters.
The velocity posterior is again double peaked because of walker build-ups at maximum velocity ($b=0$) and minimum velocity where the distribution goes flat ($b>1.0$).
Either hypothesis gives extremely fast velocities, and therefore circular periods far shorter than can be constrained by the non-detection of other events in the lightcurve.
Hence, an eclipse at the perihelion of an eccentric object (either a planet or grazing EB) is more likely.
Alternatively, the signal could simply be unexplained red noise or even an outburst of evaporation from an otherwise undetectable planet.

\subsubsection{EPIC201631267}
\texttt{Namaste} suggests this shallow but V-shaped transit is most likely a grazing eclipse, although the distributions build up at the $R_p/R_{\star}$ and $b$ limits, suggesting the true eclipsing body could be larger than our model can fit. The period of $440^{+160}_{-120}d$ is well-constrained, however, as the velocity of the eclipsing body becomes constant for extremely grazing eclipses.

\subsubsection{EPIC203914123}
The deepest and longest eclipse detected in K2, this star undergoes nearly 3 days of 2.5\% dimming.
\texttt{Namaste}, using a main-sequence approximation from photometric colours, estimates a period of 84,000d (230yr).
The length of the eclipse means fitting was extremely precise and the posterior distributions are extremely well-constrained, although some red noise (seen in the form of a V-shape during the base of the transit) remains.
However, the likelihood of seeing such an eclipse around a main sequence star is around $10^{-7}$.
Hence, our assumptions are likely wrong, and this is probably a low-mass star eclipsing a sub-giant or giant star.
Indeed, this hypothesis also explains the 'ramp' seen up to and away from the eclipse as gravity darkening or ellipsoidal variation in a tidally locked binary system.
As period is directly proportional to stellar density, a main sequence star passing in front of a $10R_{\odot}$, stellar mass giant would produce a similar eclipse on only an 80d, rather than 80,000d, orbit.
Such an orbit would be easily distiguished by RV follow-up.
Asteroseismology could be used with the K2 lightcurve to achieve better density (and therefore period) constraints for giant star eclipses such as this.
Such discoveries are important finds on their own, allowing models of stellar evolution to be tested \citep{gaulme2013red}.

\subsubsection{EPIC201720401}
Similar to EPIC203914123, \texttt{Namaste} fits this eclipse extremely well but gives an unfeasibly long period for EPIC201720401 of $6300 ^{+2200}_{-1500} d$.
Hence a giant/M-dwarf eclipsing binary is the more likely scenario, but we cannot rule out either scenario without further follow-up, such as radial velocities.
With good estimates of stellar density, however, the period could be estimated accurately and subsequent eclipses followed-up.

\subsubsection{EPIC201635132}
Similarly deep to the previous eclipses, this eclipse is also short and V-shaped, which \texttt{Namaste} identifies as a grazing eclipse with minimum impact parameter of 0.85 and minimum radius of $0.17R_{\star}$ constrained only by our imposed limits on $b$ and $R_p/R_{\star}$.
With a minimum period of 63 days from the lightcurve, the velocity is likely to be around half of the expected value, suggesting a much more grazing eclipse than possible to detect with \texttt{Namaste}.

\subsection{Source of Uncertainty}
For our best-constrained K2 candidates, the dominant source of error is stellar density.
Even for a main sequence star with a well-constrained stellar temperature, density uncertainties can be on the order of 50\%.

For less well-constrained probability distributions, for example from noisy lightcurves, the uncertainty in planetary velocity can be important.
This is especially true for small planets for which impact parameter becomes more degenerate.

As shown in section 2.4, eccentricity also has a role in increasing period uncertainties.
However, the majority of exoplanets are expected to be on circular or near-circular orbits for which the correction is minimal \citep{kipping2013parametrizing}.
If density can be constrained by follow-up observations, however, eccentricity could become an important factor when searching for a subsequent transit.

One interesting exception may be known multiplanet systems, which improve the single transit method in three ways:
The low mutual inclination in such systems increases the transit probability of exterior planets; stability constraints and formation pathways also limit long-period planets to more circular orbits; and stellar parameters such as density are significantly improved due to the information gathered by interior planets (for example densities through transit durations).
As such, searches for single transits in known multi-planet systems could have valuable results.

\subsection{Potential for Follow-up Observations}
Single spectra of candidates could immediately rule out large binary companions from cross-correlation functions.
Spectral fitting would also allow better estimates of stellar parameters, including density from log-g.
This would also rule out giant stars from the analysis.
Lucky or AO imaging could rule out close companions that could either be false positives, or diluting the transit depth.

Estimated planetary radius and orbital period can give likely RV amplitudes and determine the observability of individual targets.
Radial Velocities can be tailored to the likely period to give best observing strategy.
For objects with poorly-constrained parameters (e.g. grazing transits), RVs could be used to reduce the uncertainty on planet-to-star radius ratios.
This could then be used as a prior to re-calculate the planetary velocity, and hence improve the period estimate.


Good period estimates from single transit candidates could then be used to search for repeat transits.
Such analysis would ideally utilise observers at different longitudes to cover a likely transit range to be many days.
Long period (and low impact parameter) planets may have a better probability of re-observation due to their longer transit duration.
Precise transit survey telescopes such as MEarth \citep{irwin2008mearth} or NGTS \citep{wheatley2013next}, in combination with amateur observer programs, could be used most effectively to this goal.

For the smallest transit depths, small space-based telescopes such as MOST \citep{rucinski2003most} or Cheops \citep{broeg2013cheops} may be the only method of confirming the orbital period.

\subsection{Validation}

We have shown that, for favourable single transits, the orbit and size of a transiting exoplanet can be accurately determined, especially for giant planets. 
However, alternative sources for such transits (e.g. from background eclipsing binaries) remain.
As has been shown \citep{morton2012efficient}, the source of such false positives can be probabilistically excluded with follow-up data.
For example sub-arcsecond imaging can rule out stellar companions that could be producing spurious signals.
Spectral fitting to high-SNR spectra can be used to rule out closer binary companions.
Radial velocity measurements (even where the detection of the signal from planet is impossible) could be used to place limits on the size of any companion.
In those cases without close companions, the planetary radius of the eclipsing object can be probabilistically limited to a planetary, rather than stellar, origin.
This could, in a similar way to the validation of other single transiting systems such as Kepler-452 \citep{jenkins2015discovery}, constitute a probabilistic validation of the planet without observing subsequent transits.

\subsection{Application to Future Missions}
Single transits could be used by many future surveys to detect long-period transiting planets around bright stars.

The Transiting Exoplanet Survey Satellite (TESS,\cite{ricker2015transiting}) will monitor 200,000 stars on 2min cadence over 28d observing windows.
Studies of TESS planet yield suggest more than 100 single transits could be detected above a noise threshold of $7.3\sigma$ \citep{sullivan2015transiting}.
\texttt{Namaste} could be an important tool in the follow-up of these planets.

Initial observing plans for the PLAnetary Transits and Osscillation of Stars (PLATO) mission suggest 6 fields could be observed on 2-5 month campaigns, yielding 60,000 bright stars ($V\textrm{mag}<12$) with 30s cadence and hundreds of thousands of fainter stars ($12<V\textrm{mag}<16$) with 10-minute cadence.
The potential combination of asteroseismology-derived densities (accurate to 10\%) with high-cadence, high-precision photometric data could produce dozens of validated long-period planets \citep{osborn2015plato}.

Ground-based surveys such as MEarth \citep{irwin2008mearth} already search for the single transits of stars with an innovative detection strategy \citep{nutzman2008design}.
This method could also be applied to more traditional single-field photometric surveys such as NGTS \citep{wheatley2013next}.


Gaia has the capability to detect tens of thousands of giant planets on 1 to 4 year orbits \citep{dzigan2012detection}, hence any transiting gas giants on these orbits found by TESS, PLATO or from the ground stand a good chance of being confirmed by Gaia astrometry. 
Hence this method could lead to the first overlap between the realms of transiting and astrometric exoplanet astronomy.

The long transit duration of planets detected from a single transit also makes them amenable to transmission spectroscopy, for example by JWST \citep{belu2011primary}.
With a dearth of transiting warm and cold Jupiters known around bright stars, such objects could prove a vital link between the atmospheres of hot Jupiters and those of solar System gas giants.

\section{Conclusion}
We have developed \texttt{Namaste}, a method of combining stellar parameters with the lightcurve of a single transit to estimate orbital parameters.
We have tested this analysis on published transit parameters for a large sample of \textit{Kepler} multiplanets, showing close agreement.
A test of the full fitting method on the lightcurves of four known \textit{Kepler} planets and two KOI candidates showed extremely good agreement, with the Periods of Kepler-51 d, Kepler-117 c, Kepler-111 c and KOI1431.01 all estimated to within 10\%.

We performed an iterative search on three campaigns of K2 data and identified 7 preliminary single transit events. 
One of these, EPIC203311200, is an extremely good planet candidate with a \texttt{Namaste}-estimated period of $540 ^{+410}_{-230}$ days and a size of $0.51 \pm 0.05 R_{\textrm{Jup}}$.
We also detect three single eclipses from ambiguous but potentially planetary bodies, and three from likely eclipsing binaries.
For all candidates, future follow-up campaigns are vital to determine the source of the eclipse and better constrain the orbital parameters.

\texttt{Namaste} is therefore a useful tool, allowing more targeted follow-up observations with radial velocity measurements and the search for additional transits.
For giant planets around bright stars, subsequent RV follow-up could lead to their confirmation.
Smaller planets around fainter stars could still be validated with follow-up techniques such as diffraction-limited imaging and high-resolution spectroscopy.

In future the detection and analysis of single transits in this way could lead K2, TESS, PLATO and ground-based photometric surveys to detect transiting exoplanets on orbits longer than their observing campaigns would traditionally allow.
Such long-period planets, especially if found around bright stars, could pave the way for a new regime of exoplanetary science. 
This includes the detection of planets within the astrometric sensitivity of Gaia, and cold Jupiters with atmospheres observable in transmission spectroscopy by JWST. 
EPIC203311200 b, if proved to be planetary by ongoing follow-up work, could fit both roles.

\section{Acknowledgements}
Thanks to Warwick's team of eyeballers, who stared at far too many lightcurves for far too little reward.

This research was funded by a Chancellors Scholarship from the University of Warwick.

Some of the data presented in this paper were obtained from the Mikulski Archive for Space Telescopes (MAST). STScI is operated by the Association of Universities for Research in Astronomy, Inc., under NASA contract NAS5-26555. Support for MAST for non-HST data is provided by the NASA Office of Space Science via grant NNX09AF08G and by other grants and contracts.

This paper includes data collected by the \textit{Kepler} mission. Funding for the \textit{Kepler} mission is provided by the NASA Science Mission directorate.




\bibliographystyle{mnras}
\bibliography{refs} 

\begin{thebibliography}{}
\makeatletter
\relax
\def\mn@urlcharsother{\let\do\@makeother \do\$\do\&\do\#\do\^\do\_\do\%\do\~}
\def\mn@doi{\begingroup\mn@urlcharsother \@ifnextchar [ {\mn@doi@}
  {\mn@doi@[]}}
\def\mn@doi@[#1]#2{\def\@tempa{#1}\ifx\@tempa\@empty \href
  {http://dx.doi.org/#2} {doi:#2}\else \href {http://dx.doi.org/#2} {#1}\fi
  \endgroup}
\def\mn@eprint#1#2{\mn@eprint@#1:#2::\@nil}
\def\mn@eprint@arXiv#1{\href {http://arxiv.org/abs/#1} {{\tt arXiv:#1}}}
\def\mn@eprint@dblp#1{\href {http://dblp.uni-trier.de/rec/bibtex/#1.xml}
  {dblp:#1}}
\def\mn@eprint@#1:#2:#3:#4\@nil{\def\@tempa {#1}\def\@tempb {#2}\def\@tempc
  {#3}\ifx \@tempc \@empty \let \@tempc \@tempb \let \@tempb \@tempa \fi \ifx
  \@tempb \@empty \def\@tempb {arXiv}\fi \@ifundefined
  {mn@eprint@\@tempb}{\@tempb:\@tempc}{\expandafter \expandafter \csname
  mn@eprint@\@tempb\endcsname \expandafter{\@tempc}}}

\bibitem[\protect\citeauthoryear{Akeson et~al.,}{Akeson
  et~al.}{2013}]{akeson2013nasa}
Akeson R.,  et~al., 2013, Publications of the Astronomical Society of the
  Pacific, 125, 989

\bibitem[\protect\citeauthoryear{Armstrong, Osborn, Brown, Faedi, Chew, Martin,
  Pollacco  \& Udry}{Armstrong et~al.}{2014}]{armstrong2014abundance}
Armstrong D.~J.,  Osborn H.,  Brown D.,  Faedi F.,  Chew Y. G.~M.,  Martin
  D.~V.,  Pollacco D.,   Udry S.,  2014, Monthly Notices of the Royal
  Astronomical Society, 444, 1873

\bibitem[\protect\citeauthoryear{Armstrong et~al.,}{Armstrong
  et~al.}{2015b}]{armstrong2015k2}
Armstrong D.,  et~al., 2015b, arXiv preprint arXiv:1502.04004

\bibitem[\protect\citeauthoryear{Armstrong et~al.,}{Armstrong
  et~al.}{2015a}]{armstrong2015one}
Armstrong D.~J.,  et~al., 2015a, arXiv preprint arXiv:1503.00692

\bibitem[\protect\citeauthoryear{Barclay et~al.,}{Barclay
  et~al.}{2013}]{barclay2013sub}
Barclay T.,  et~al., 2013, Nature, 494, 452

\bibitem[\protect\citeauthoryear{Barnes}{Barnes}{2007}]{barnes2007effects}
Barnes J.~W.,  2007, Publications of the Astronomical Society of the Pacific,
  119, 986

\bibitem[\protect\citeauthoryear{Belu, Selsis, Morales, Ribas, Cossou  \&
  Rauer}{Belu et~al.}{2011}]{belu2011primary}
Belu A.,  Selsis F.,  Morales J.-C.,  Ribas I.,  Cossou C.,   Rauer H.,  2011,
  Astronomy \& Astrophysics, 525, A83

\bibitem[\protect\citeauthoryear{Bennett \& Rhie}{Bennett \&
  Rhie}{1996}]{bennett1996detecting}
Bennett D.~P.,  Rhie S.~H.,  1996, The Astrophysical Journal, 472, 660

\bibitem[\protect\citeauthoryear{Borucki et~al.,}{Borucki
  et~al.}{2010}]{borucki2010kepler}
Borucki W.~J.,  et~al., 2010, Science, 327, 977

\bibitem[\protect\citeauthoryear{Broeg et~al.,}{Broeg
  et~al.}{2013}]{broeg2013cheops}
Broeg C.,  et~al., 2013, in EPJ Web of Conferences. p. 03005

\bibitem[\protect\citeauthoryear{Bruno et~al.,}{Bruno
  et~al.}{2015}]{bruno2015sophie}
Bruno G.,  et~al., 2015, in EPJ Web of Conferences. p. 06014

\bibitem[\protect\citeauthoryear{Collier-Cameron et~al.,}{Collier-Cameron
  et~al.}{2007}]{cameron2007efficient}
Collier-Cameron A.,  et~al., 2007, Monthly Notices of the Royal Astronomical
  Society, 380, 1230

\bibitem[\protect\citeauthoryear{Cutri et~al.,}{Cutri
  et~al.}{2003}]{cutri20032mass}
Cutri R.,  et~al., 2003

\bibitem[\protect\citeauthoryear{Dzigan \& Zucker}{Dzigan \&
  Zucker}{2012}]{dzigan2012detection}
Dzigan Y.,  Zucker S.,  2012, The Astrophysical Journal Letters, 753, L1

\bibitem[\protect\citeauthoryear{Fitzgerald}{Fitzgerald}{1970}]{fitzgerald1970intrinsic}
Fitzgerald M.,  1970, Astronomy and Astrophysics, 4, 234

\bibitem[\protect\citeauthoryear{Foreman-Mackey, Hogg, Lang  \&
  Goodman}{Foreman-Mackey et~al.}{2013}]{foreman2013emcee}
Foreman-Mackey D.,  Hogg D.~W.,  Lang D.,   Goodman J.,  2013, Publications of
  the Astronomical Society of the Pacific, 125, 306

\bibitem[\protect\citeauthoryear{Foreman-Mackey, Montet, Hogg, Morton, Wang  \&
  Sch{\"o}lkopf}{Foreman-Mackey et~al.}{2015}]{foreman2015systematic}
Foreman-Mackey D.,  Montet B.~T.,  Hogg D.~W.,  Morton T.~D.,  Wang D.,
  Sch{\"o}lkopf B.,  2015, arXiv preprint arXiv:1502.04715

\bibitem[\protect\citeauthoryear{Fressin et~al.,}{Fressin
  et~al.}{2013}]{fressin2013false}
Fressin F.,  et~al., 2013, The Astrophysical Journal, 766, 81

\bibitem[\protect\citeauthoryear{Gaulme, McKeever, Rawls, Jackiewicz, Mosser
  \& Guzik}{Gaulme et~al.}{2013}]{gaulme2013red}
Gaulme P.,  McKeever J.,  Rawls M.~L.,  Jackiewicz J.,  Mosser B.,   Guzik J.,
  2013, The Astrophysical Journal, 767, 82

\bibitem[\protect\citeauthoryear{H{\o}g et~al.,}{H{\o}g
  et~al.}{2000}]{hog2000tycho}
H{\o}g E.,  et~al., 2000, Astronomy and Astrophysics, 355, L27

\bibitem[\protect\citeauthoryear{Howell et~al.,}{Howell
  et~al.}{2014}]{howell2014k2}
Howell S.~B.,  et~al., 2014, Publications of the Astronomical Society of the
  Pacific, 126, 398

\bibitem[\protect\citeauthoryear{Irwin, Charbonneau, Nutzman  \& Falco}{Irwin
  et~al.}{2008}]{irwin2008mearth}
Irwin J.,  Charbonneau D.,  Nutzman P.,   Falco E.,  2008, Proceedings of the
  International Astronomical Union, 4, 37

\bibitem[\protect\citeauthoryear{Jenkins et~al.,}{Jenkins
  et~al.}{2015}]{jenkins2015discovery}
Jenkins J.~M.,  et~al., 2015, The Astronomical Journal, 150, 56

\bibitem[\protect\citeauthoryear{Jones, Oliphant, Peterson  et~al.}{Jones
  et~al.}{01  }]{jones2001}
Jones E.,  Oliphant T.,  Peterson P.,   et~al., 2001--, {SciPy}: Open source
  scientific tools for {Python}, \url {http://www.scipy.org/}

\bibitem[\protect\citeauthoryear{Jontof-Hutter, Lissauer, Rowe  \&
  Fabrycky}{Jontof-Hutter et~al.}{2014}]{jontof2014kepler}
Jontof-Hutter D.,  Lissauer J.~J.,  Rowe J.~F.,   Fabrycky D.~C.,  2014, The
  Astrophysical Journal, 785, 15

\bibitem[\protect\citeauthoryear{Kipping}{Kipping}{2013}]{kipping2013parametrizing}
Kipping D.~M.,  2013, Monthly Notices of the Royal Astronomical Society:
  Letters, 434, L51

\bibitem[\protect\citeauthoryear{Kipping et~al.,}{Kipping
  et~al.}{2014}]{kipping2014discovery}
Kipping D.~M.,  et~al., 2014, The Astrophysical Journal, 795, 25

\bibitem[\protect\citeauthoryear{Mandel \& Agol}{Mandel \&
  Agol}{2002}]{mandel2002analytic}
Mandel K.,  Agol E.,  2002, The Astrophysical Journal Letters, 580, L171

\bibitem[\protect\citeauthoryear{Masuda}{Masuda}{2014}]{masuda2014very}
Masuda K.,  2014, The Astrophysical Journal, 783, 53

\bibitem[\protect\citeauthoryear{Mayor et~al.,}{Mayor
  et~al.}{2011}]{mayor2011harps}
Mayor M.,  et~al., 2011, arXiv preprint arXiv:1109.2497

\bibitem[\protect\citeauthoryear{Montet et~al.,}{Montet
  et~al.}{2015}]{montet2015stellar}
Montet B.~T.,  et~al., 2015, The Astrophysical Journal, 809, 25

\bibitem[\protect\citeauthoryear{Morton}{Morton}{2012}]{morton2012efficient}
Morton T.~D.,  2012, The Astrophysical Journal, 761, 6

\bibitem[\protect\citeauthoryear{Nutzman \& Charbonneau}{Nutzman \&
  Charbonneau}{2008}]{nutzman2008design}
Nutzman P.,  Charbonneau D.,  2008, Publications of the Astronomical Society of
  the Pacific, 120, 317

\bibitem[\protect\citeauthoryear{Osborn, Brown  \& Pollacco}{Osborn
  et~al.}{prep}]{osborn2015plato}
Osborn H.~P.,  Brown D.~J.,   Pollacco D.,  in prep,

\bibitem[\protect\citeauthoryear{Price, Rogers, Johnson  \& Dawson}{Price
  et~al.}{2015}]{price2015low}
Price E.~M.,  Rogers L.~A.,  Johnson J.~A.,   Dawson R.~I.,  2015, The
  Astrophysical Journal, 799, 17

\bibitem[\protect\citeauthoryear{Ricker et~al.,}{Ricker
  et~al.}{2015}]{ricker2015transiting}
Ricker G.~R.,  et~al., 2015, Journal of Astronomical Telescopes, Instruments,
  and Systems, 1, 014003

\bibitem[\protect\citeauthoryear{Rowe et~al.,}{Rowe
  et~al.}{2014}]{rowe2014validation}
Rowe J.~F.,  et~al., 2014, The Astrophysical Journal, 784, 45

\bibitem[\protect\citeauthoryear{Rucinski, Carroll, Kuschnig, Matthews  \&
  Stibrany}{Rucinski et~al.}{2003}]{rucinski2003most}
Rucinski S.,  Carroll K.,  Kuschnig R.,  Matthews J.,   Stibrany P.,  2003,
  Advances in Space Research, 31, 371

\bibitem[\protect\citeauthoryear{Seager \& Mallen-Ornelas}{Seager \&
  Mallen-Ornelas}{2003}]{seager2003unique}
Seager S.,  Mallen-Ornelas G.,  2003, The Astrophysical Journal, 585, 1038

\bibitem[\protect\citeauthoryear{Sing}{Sing}{2010}]{sing2010stellar}
Sing D.~K.,  2010, Astronomy \& Astrophysics, 510, A21

\bibitem[\protect\citeauthoryear{Sullivan et~al.,}{Sullivan
  et~al.}{2015}]{sullivan2015transiting}
Sullivan P.~W.,  et~al., 2015, The Astrophysical Journal, 809, 77

\bibitem[\protect\citeauthoryear{Torres, Andersen  \& Gim{\'e}nez}{Torres
  et~al.}{2010}]{torres2010accurate}
Torres G.,  Andersen J.,   Gim{\'e}nez A.,  2010, The Astronomy and
  Astrophysics Review, 18, 67

\bibitem[\protect\citeauthoryear{Van~Eylen et~al.,}{Van~Eylen
  et~al.}{2014}]{van2014asteroseismology}
Van~Eylen V.,  et~al., 2014, The Astrophysical Journal, 782, 14

\bibitem[\protect\citeauthoryear{Vanderburg \& Johnson}{Vanderburg \&
  Johnson}{2014}]{vanderburg2014technique}
Vanderburg A.,  Johnson J.~A.,  2014, Publications of the Astronomical Society
  of the Pacific, 126, 948

\bibitem[\protect\citeauthoryear{Vanderburg et~al.,}{Vanderburg
  et~al.}{2015}]{vanderburg2015characterizing}
Vanderburg A.,  et~al., 2015, The Astrophysical Journal, 800, 59

\bibitem[\protect\citeauthoryear{Wang et~al.,}{Wang
  et~al.}{2013}]{wang2013planet}
Wang J.,  et~al., 2013, The Astrophysical Journal, 776, 10

\bibitem[\protect\citeauthoryear{Wheatley et~al.,}{Wheatley
  et~al.}{2013}]{wheatley2013next}
Wheatley P.~J.,  et~al., 2013, in EPJ Web of Conferences. p. 13002

\makeatother
\end{thebibliography}



\appendix
\section{MCMC Posterior Distributions}

\begin{figure}
	\includegraphics[width=\columnwidth]{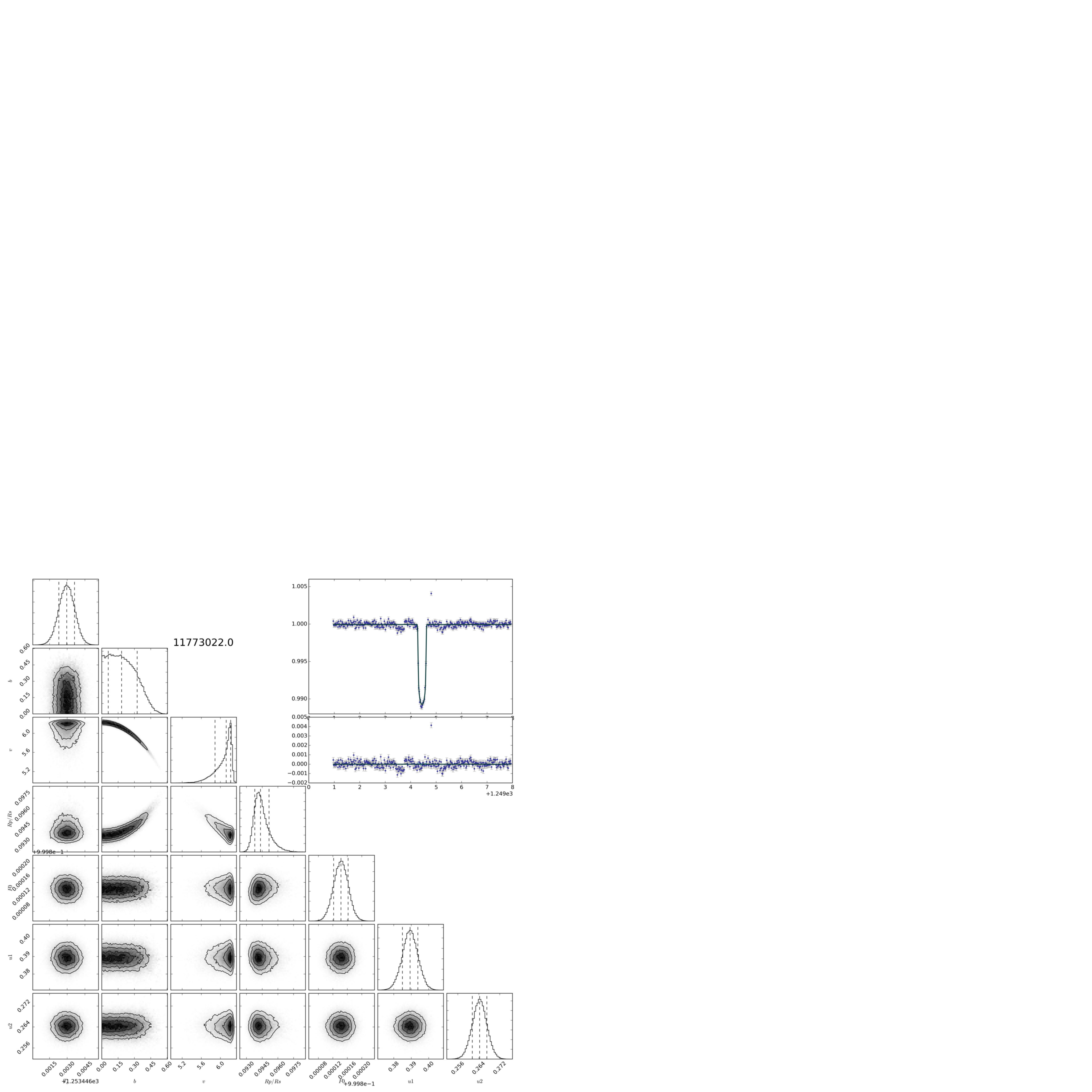}
    \caption{Posterior distribution for Kepler-51 d (KIC11773022).}
    \label{fig:Kep51}
\end{figure}
\begin{figure}
	\includegraphics[width=\columnwidth]{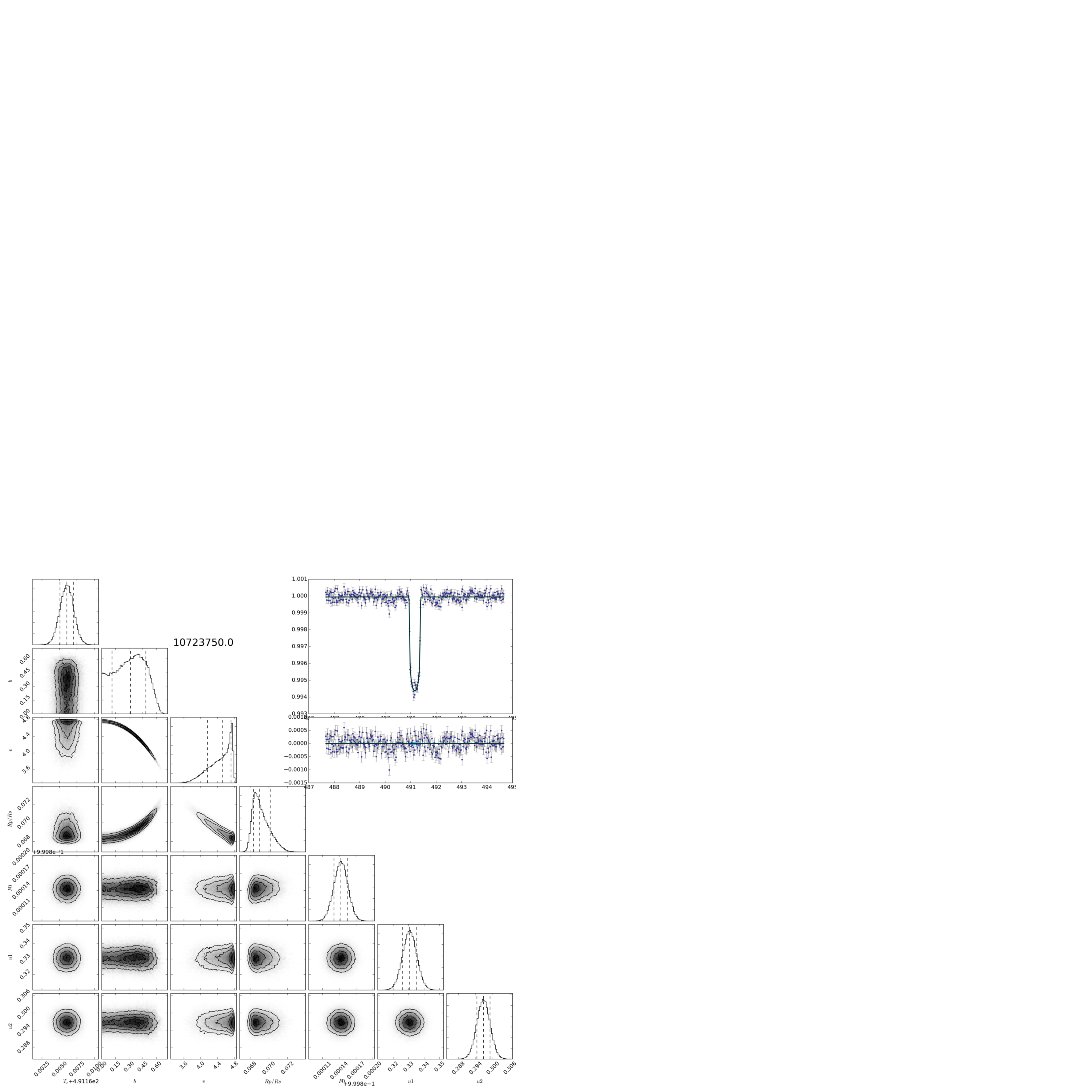}
    \caption{Posterior distribution for Kepler-117 c (KIC10723750).}
    \label{fig:Kep117}
\end{figure}
\begin{figure}
	\includegraphics[width=\columnwidth]{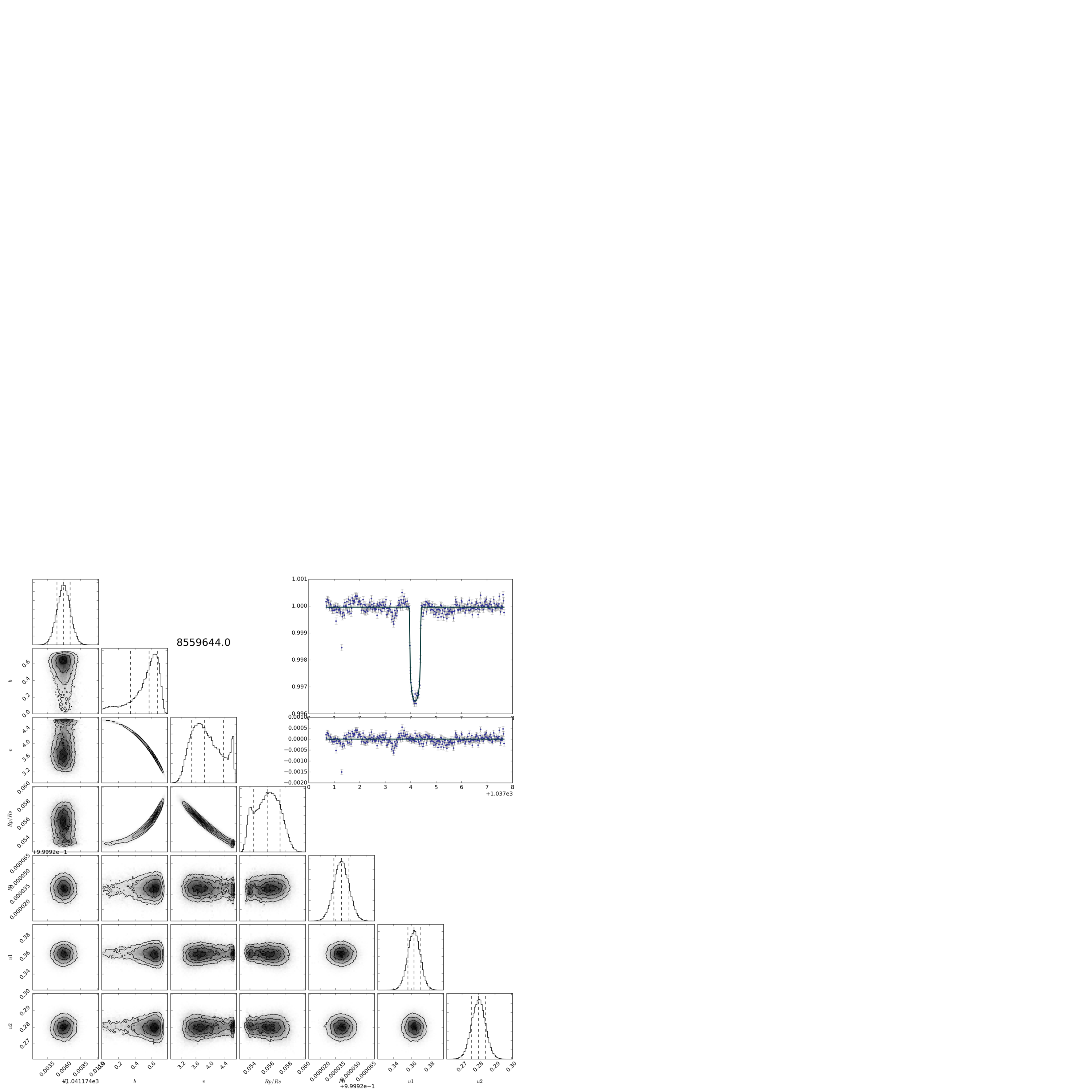}
    \caption{Posterior distribution for Kepler-111 c (KIC8559644).}
    \label{fig:Kep111}
\end{figure}
\begin{figure}
	\includegraphics[width=\columnwidth]{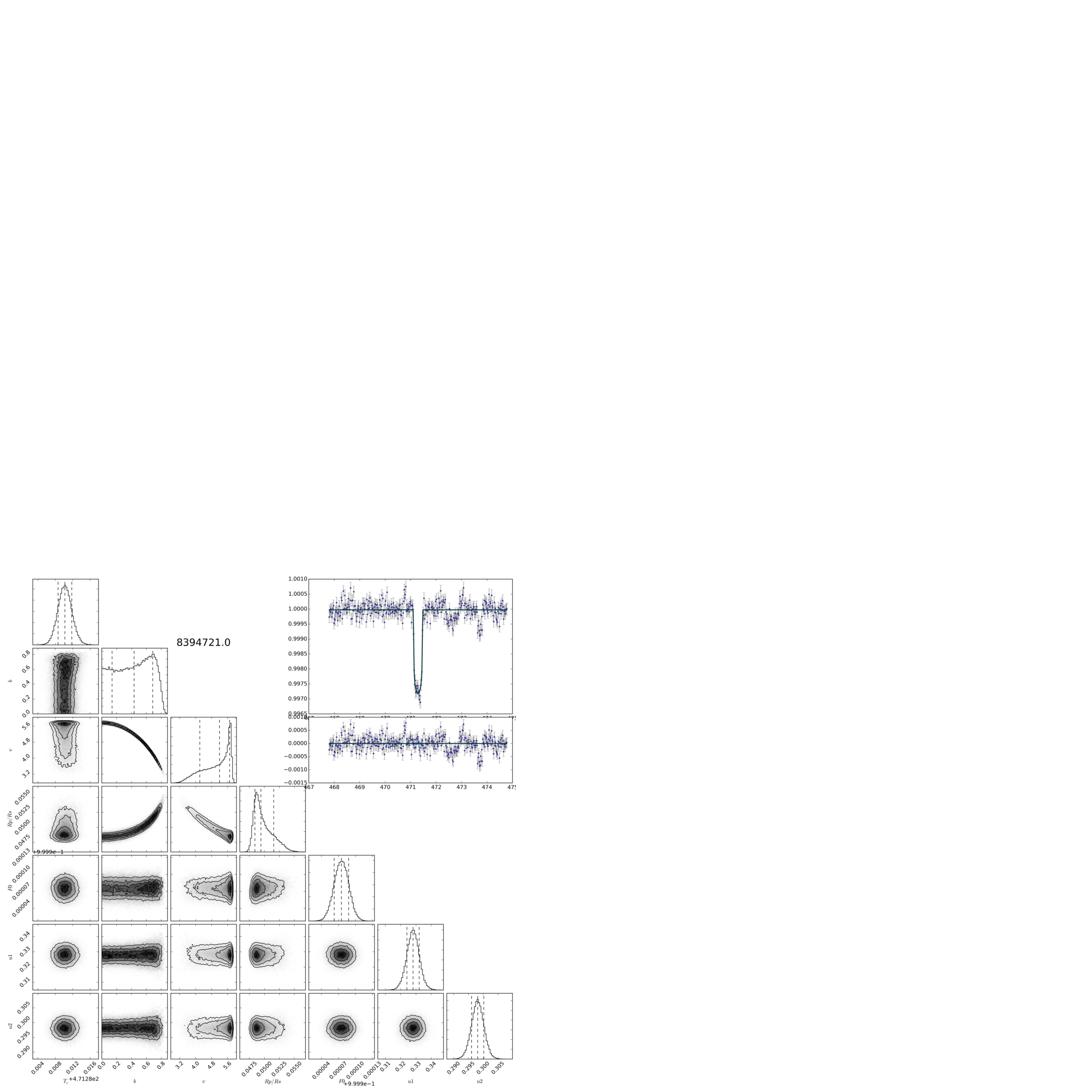}
    \caption{Posterior distribution for Kepler-79 d (KIC8394721).}
    \label{fig:Kep79}
\end{figure}
\begin{figure}
	\includegraphics[width=\columnwidth]{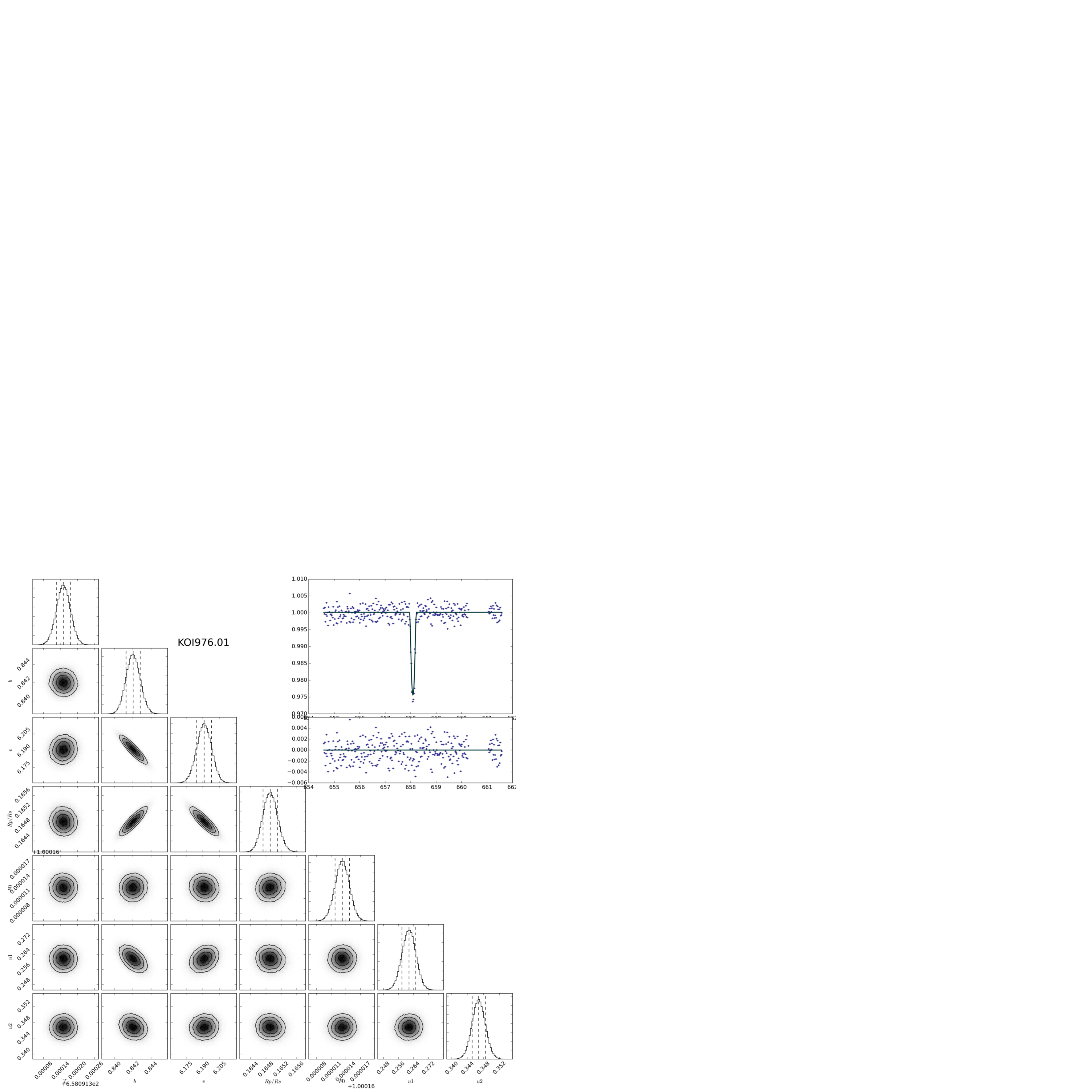}
    \caption{Posterior distribution for KOI 976.01 (KIC34417842).}
    \label{fig:KOI976}
\end{figure}

\begin{figure}
	\includegraphics[width=\columnwidth]{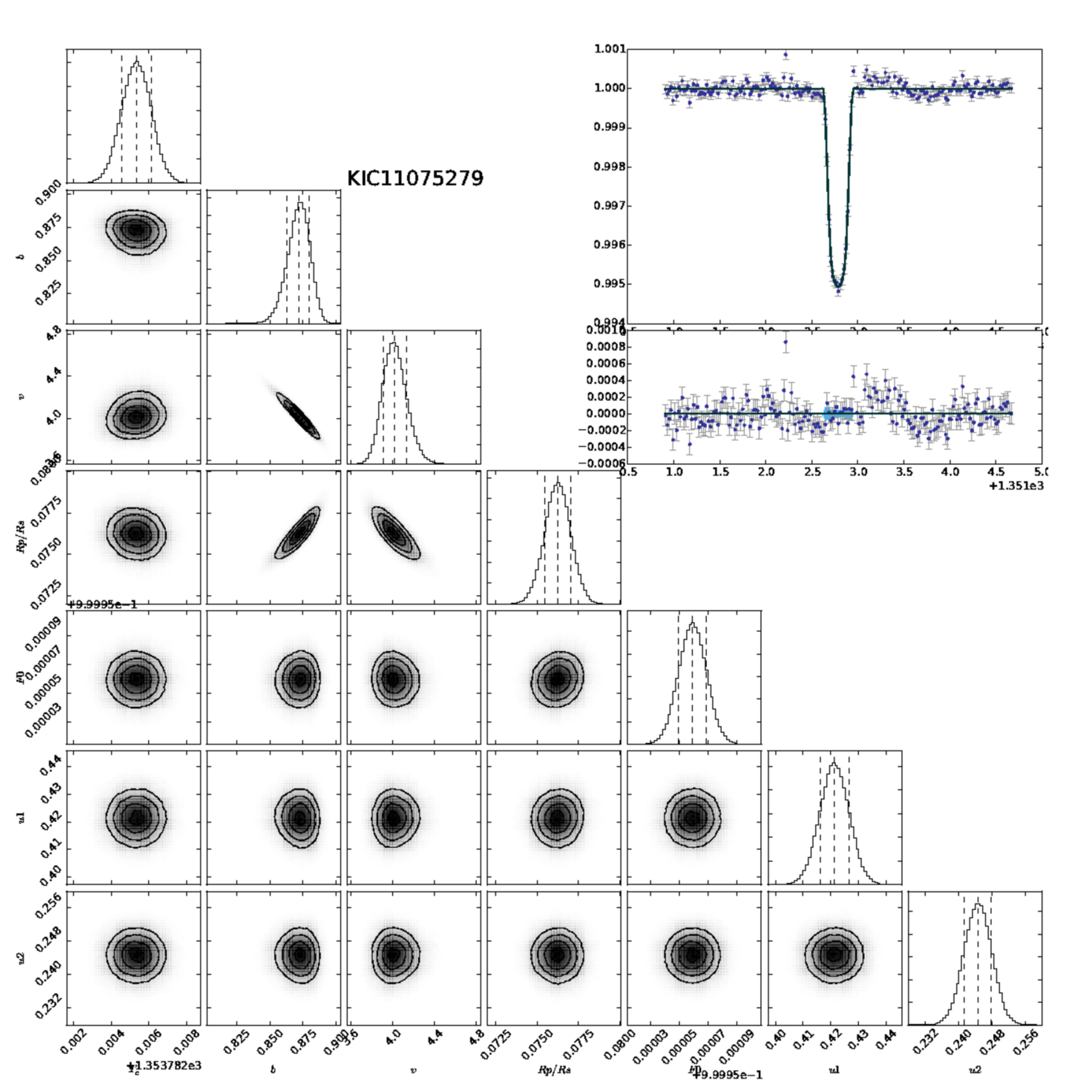}
    \caption{Posterior distribution for KOI 1431.01 (KIC11075279).}
    \label{fig:KOI1431}
\end{figure}

\begin{figure}
	\includegraphics[width=\columnwidth]{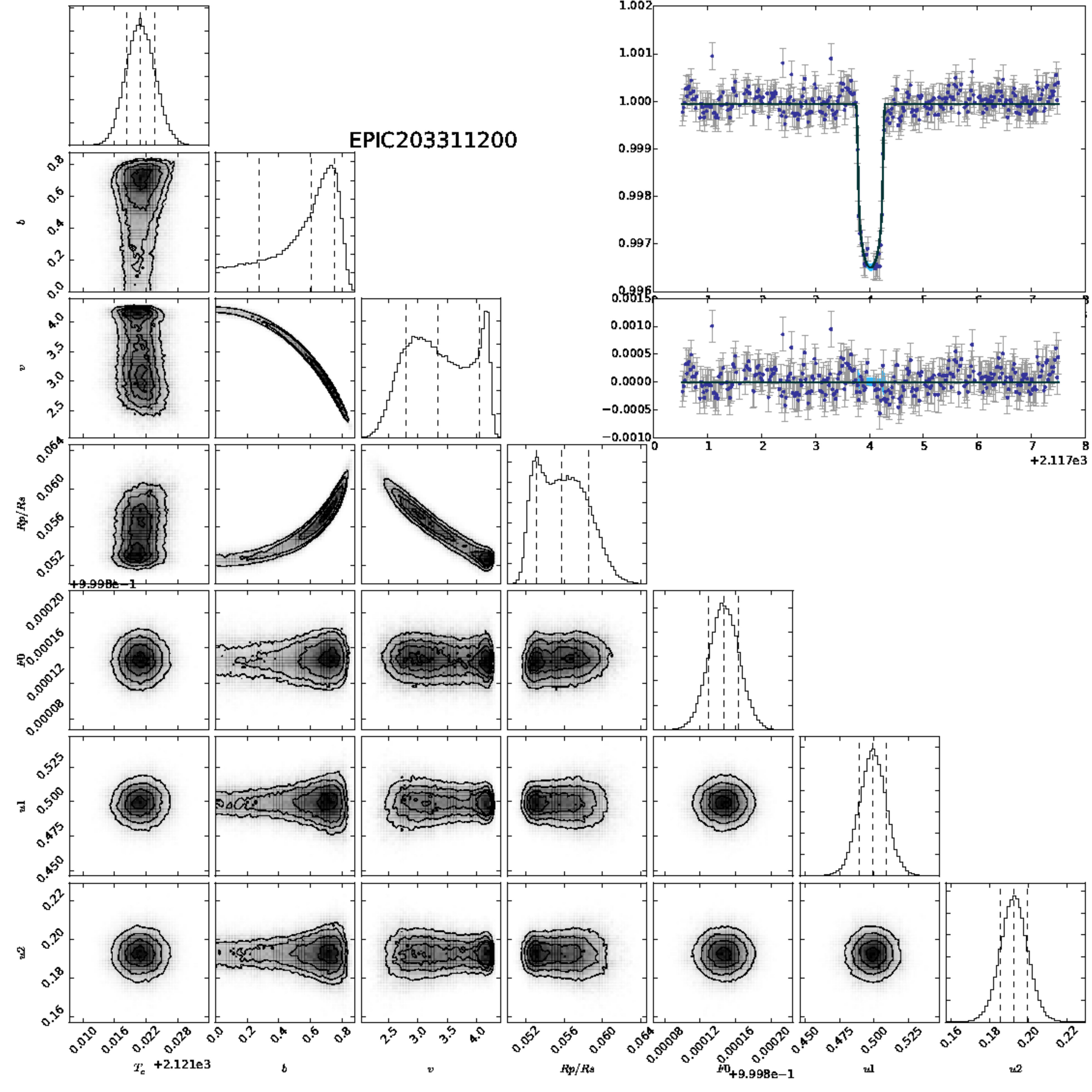}
    \caption{Posterior distribution for EPIC203311200.}
    \label{fig:EPIC203311200}
\end{figure}

\begin{figure}
	\includegraphics[width=\columnwidth]{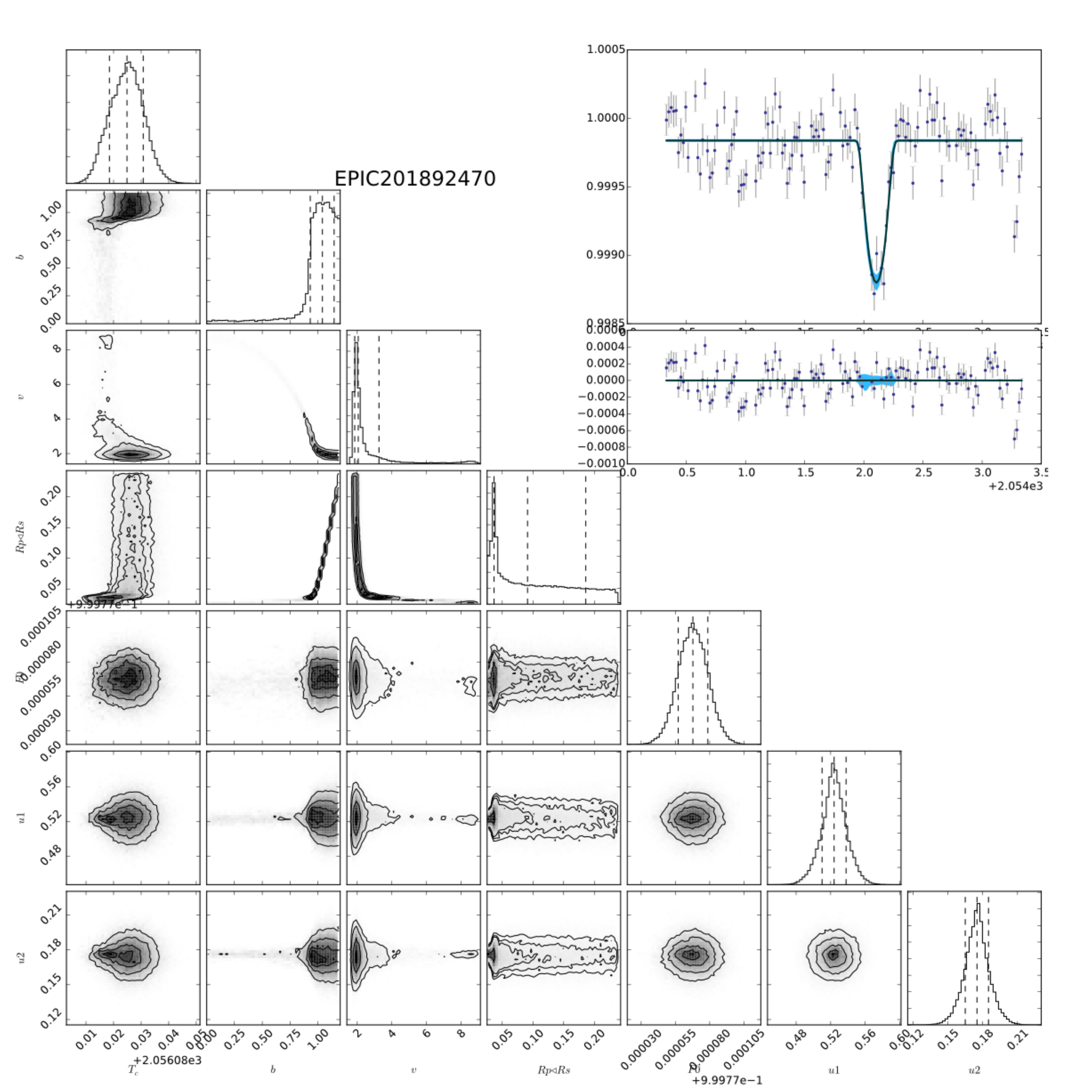}
    \caption{Posterior distribution for EPIC201892470).}
    \label{fig:EPIC201892470}
\end{figure}

\begin{figure}
	\includegraphics[width=\columnwidth]{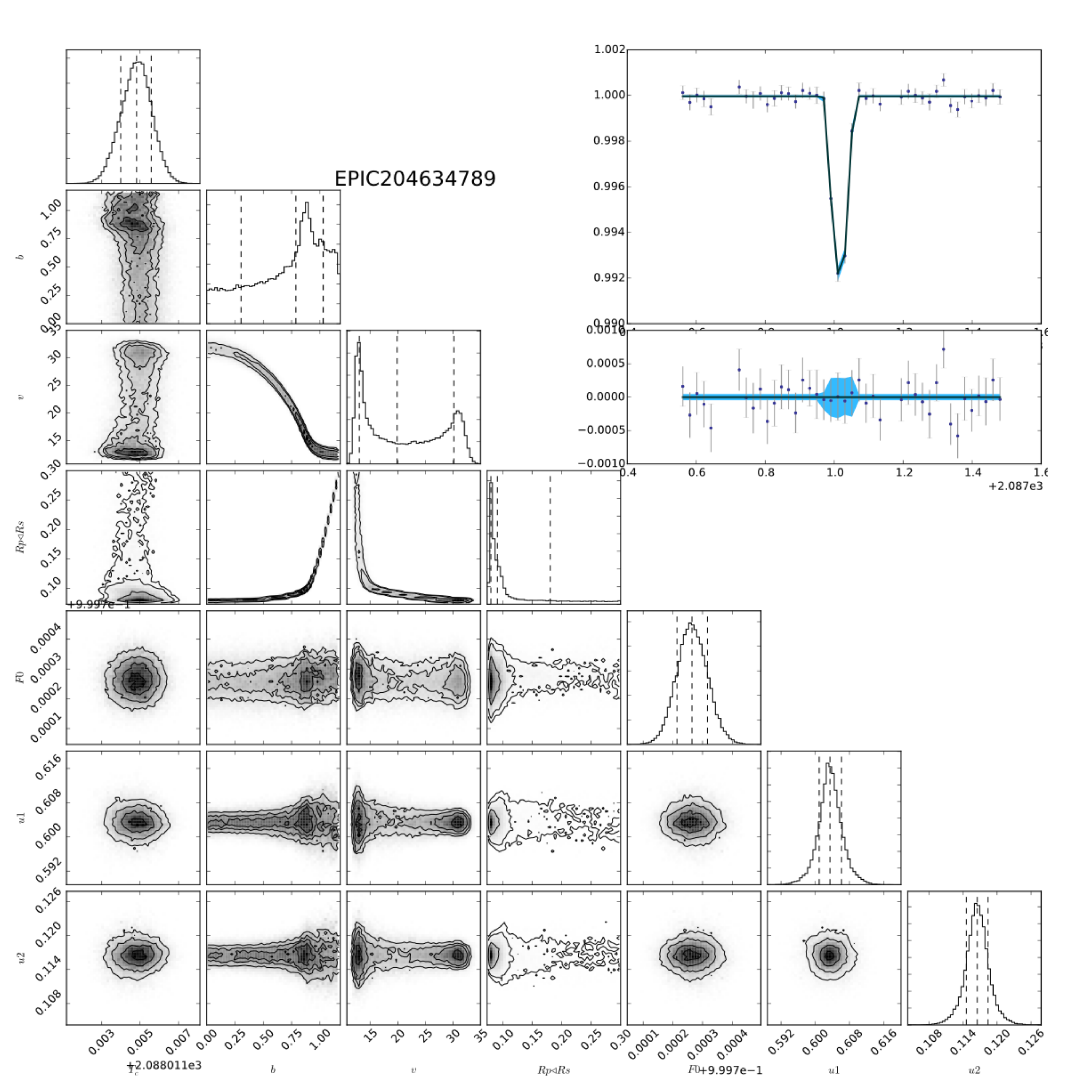}
    \caption{Posterior distribution for EPIC204634789).}
    \label{fig:EPIC204634789}
\end{figure}

\begin{figure}
	\includegraphics[width=\columnwidth]{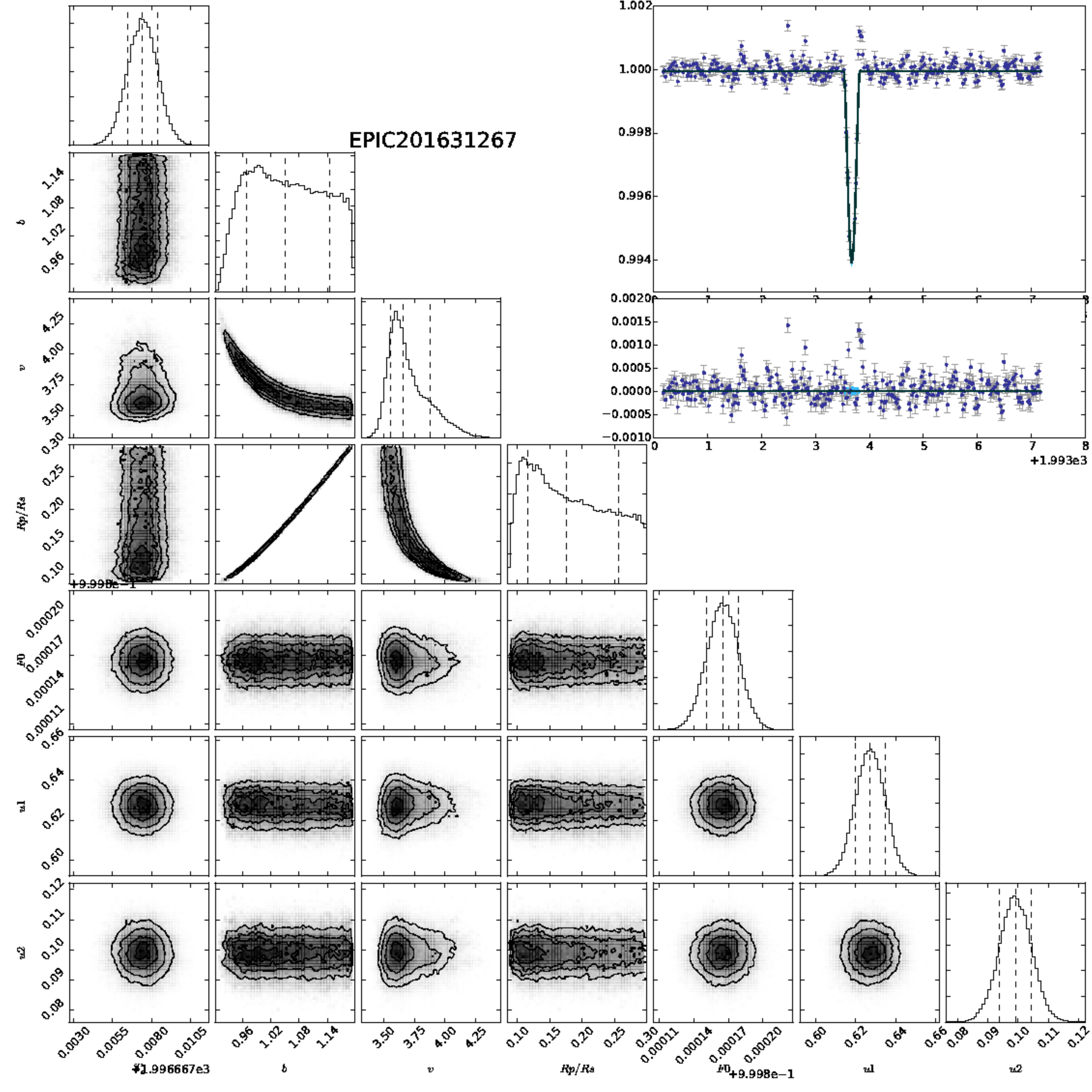}
    \caption{Posterior distribution for EPIC201631267).}
    \label{fig:EPIC201631267}
\end{figure}

\begin{figure}
	\includegraphics[width=\columnwidth]{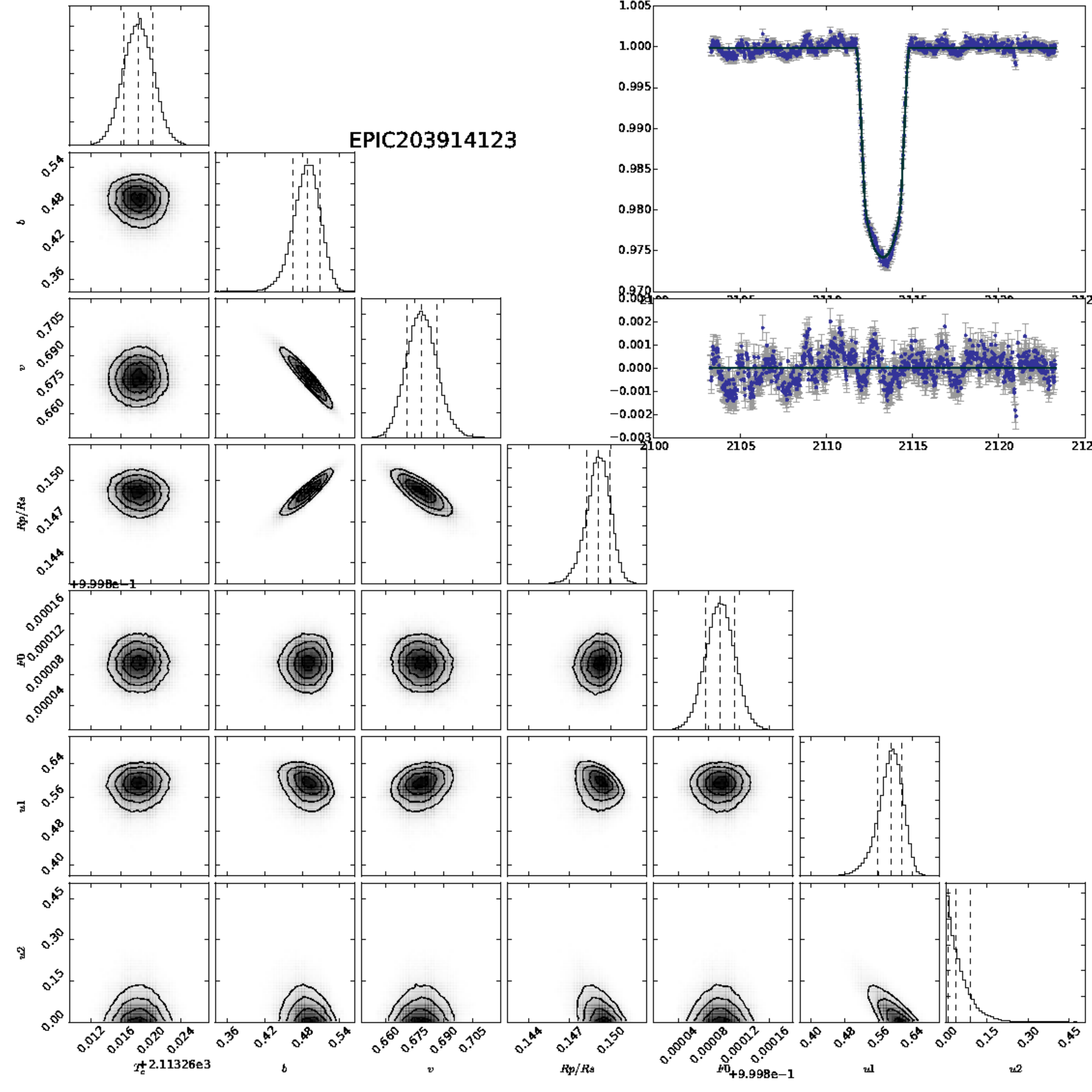}
    \caption{Posterior distribution for EPIC203914123).}
    \label{fig:EPIC203914123}
\end{figure}

\begin{figure}
	\includegraphics[width=\columnwidth]{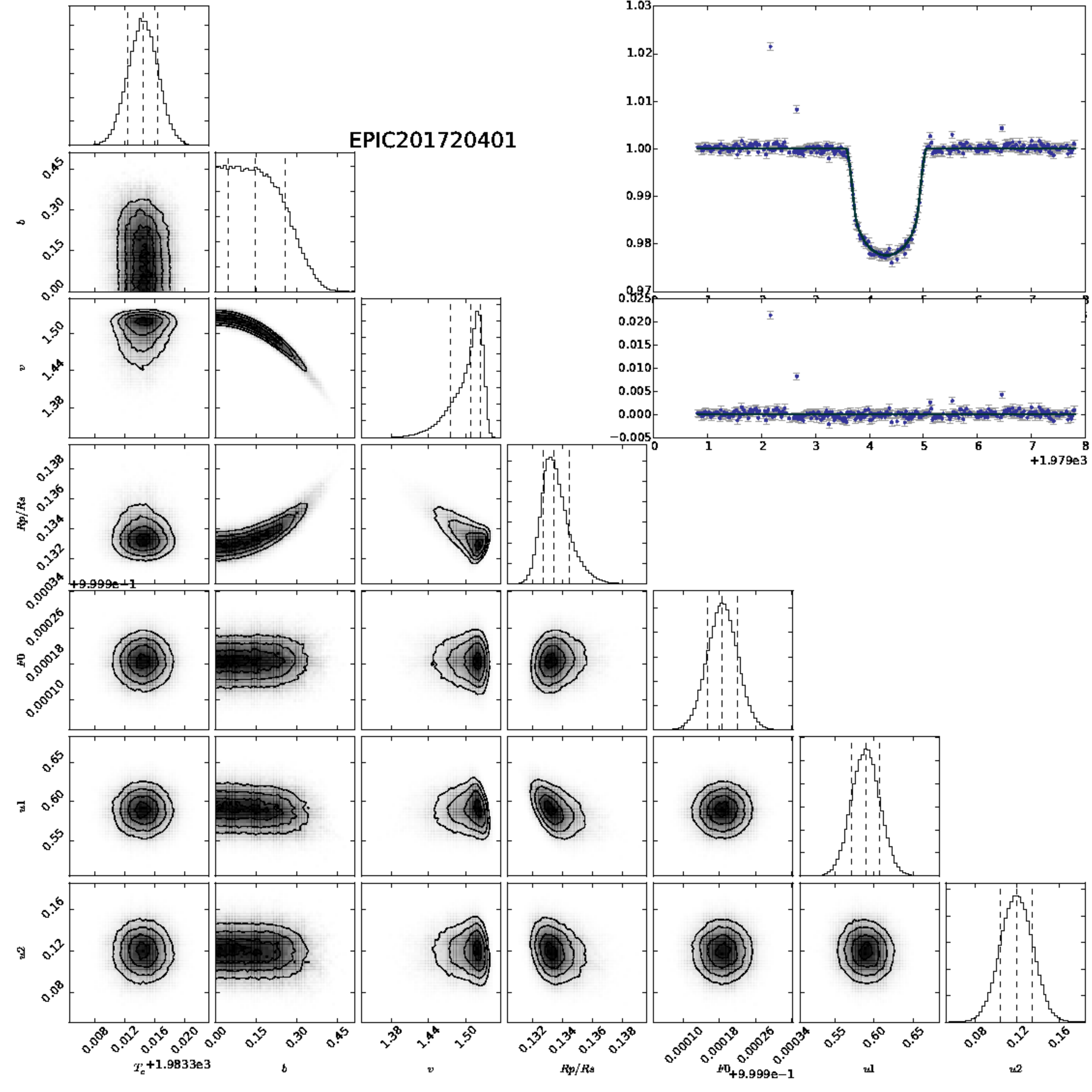}
    \caption{Posterior distribution for EPIC201720401).}
    \label{fig:EPIC201720401}
\end{figure}

\begin{figure}
	\includegraphics[width=\columnwidth]{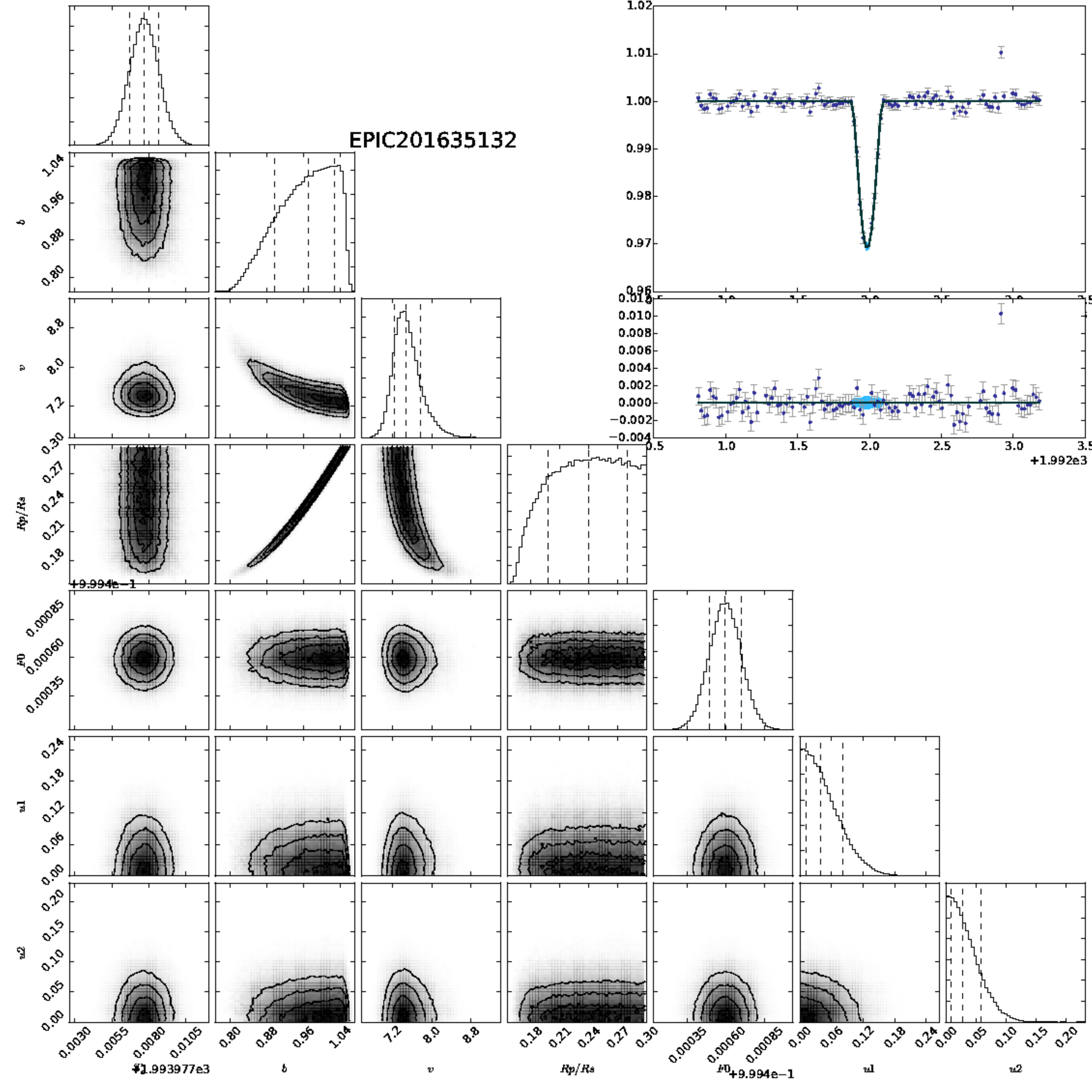}
    \caption{Posterior distribution for EPIC201635132).}
    \label{fig:EPIC201635132}
\end{figure}



\bsp	
\label{lastpage}
\end{document}